\RequirePackage{fix-cm}
\documentclass[smallextended]{svjour3}       % onecolumn (second format)
\smartqed  % flush right qed marks, e.g. at end of proof
\usepackage{graphicx}
\usepackage{mathpazo}
\usepackage{algorithm}
\usepackage{algorithmic}
\usepackage{array}

\usepackage{booktabs}
\usepackage{subcaption}
\usepackage{multirow}
\usepackage{longtable}
\usepackage[colorlinks,bookmarksopen,bookmarksnumbered,citecolor=red,urlcolor=red]{hyperref}
\usepackage{epstopdf}
\usepackage{mathpazo}
\usepackage{amsfonts}
\usepackage{amssymb}
\usepackage{mathrsfs}
\usepackage{amsmath}
\usepackage{hyperref}
\usepackage{bookmark}
\usepackage{times}
\usepackage{float}

\begin{document}

\title{Energy-efficient Deployment of Relay Nodes in Wireless Sensor Networks using Evolutionary Techniques\\
%\thanks{Grants or other notes+
\newline
}



\author{Babajide ~O.~Ayinde \and Hashim ~A.~Hashim}


\institute{B.~O.~Ayinde \at Department of Electrical and Computer Engineering, University of Louisville, Louisville, KY, 40203.
              \email{babajide.ayinde@louisville.edu.}
           %  \\
           \and
          H.~A.~Hashim \at
          Department of Electrical and Computer Engineering, The University of Western Ontario, London, ON, Canada, N6A-5B9.
             \email{hmoham33@uwo.ca}          %  \\
          %  \\
}


\maketitle

\begin{abstract}
Random deployment of sensor nodes is susceptible to initial communication hole, even when the network is densely populated. However, eliminating holes using structural deployment poses its difficulties. In either case, the resulting coverage holes can degrade overall network performance and lifetime. Many solutions utilizing Relay Nodes (RNs) have been proposed to alleviate this problem. In this regard, one of the recent solutions proposed using Artificial Bee Colony (ABC) to deploy RNs. This paper proposes RN deployment using two other evolutionary techniques - Gravitational Search Algorithm (GSA) and Differential Evolution (DE) and compares them with existing solution that uses ABC. These popular optimization tools are deployed to optimize the positions of relay nodes for lifetime maximization. Proposed algorithms guarantee satisfactory RNs utilization while maintaining desired connectivity level. It is shown that DE-based deployment improves the network lifetime better than other optimization heuristics considered.
\keywords{Artificial Bee Colony\and Gravitational Search Algorithm\and Differential Evolution\and Constrained Optimization\and Relay Deployment\and Laplacian\and Lifetime Enhancement\and Graph-based Deployment.}
\end{abstract}

\textbf{Acronyms:}\\
    \begin{tabular}{lcl}
    ABC &:& Artificial Bee Colony.\\
    CH &:& Cluster Head.\\
    DE &:& Differential Evolution.\\
    FPRNs &:& First-phase Relay Nodes.\\
    GSA &:& Gravitational Search Algorithm.\\
    MAC &:& Media Access Control.\\
    MST &:& Minimum Spanning Tree.\\
    NP-hard &:& Non-deterministic Polynomial-time Hard.\\
    O3DwLC &:& Optimized 3-D deployment with Lifetime Constraints.\\
    RNs &:& Relay Nodes.\\
    SPRN &:& Second-phase Relay Nodes.\\
    WSNs &:& Wireless Sensor Networks.\\
    \end{tabular}

\begin{table}[!hbp]\caption{Table of Notations}
\begin{center}% used the environment to augment the vertical space
% between the caption and the table
\begin{tabular}{r c p{6cm}}
\toprule
$E_p$ &:& Total energy consumed by a node\\
$I_R$ &:& Initial number of rounds using FPRN\\
$T_R$ &:& Network lifetime\\
$\lambda_2$ &:& Fielder value\\
$L$ &:& Laplacian matrix\\
$A$ &:& Incidence matrix\\
$G$ &:& Graph\\
$B$ &:& Network backbone\\
$W$ &:& Wiener index\\
$\mu$ &:& Average inter-node distance\\
$\beta$ &:& Penalty factor\\
$MP$ &:& Mutation probability.\\
$CP$ &:& Crossover probability.\\
$x$ &:& Position of particle or solution.\\
$v$ &:& Velocity of particle or solution.\\
$a$ &:& Particle acceleration.\\
$N$ &:& Number of particles or solutions in one generation.\\
$G_c$ &:& Gravitational constant.\\
$M$ &:& Mass of agent.\\
$F$ &:& Gravitational force.\\
$R$ &:& Euclidean distance.\\
\bottomrule
\end{tabular}
\end{center}
\label{TableNotations}
\end{table}

\section{Introduction}
\indent
Application domains such as environmental monitoring and military field surveillance have witnessed an increasing interest in wireless sensor networks (WSNs). In most of these applications, wireless sensors are deployed and expected to capture and then transmit parameters of interest such as temperature, pressure, and chemical activity to remote locations \cite{younis2006node,werner2005monitoring}. WSN nodes are tiny, low-powered, and multi-functional and through collaboration they are able to perform complex tasks. From design viewpoint, optimizing the connectivity, lifetime, and cost of WSNs simultaneously is generally considered an intricate task \cite{Turjman:2012,Chang:2012}. Dense deployment and unattended nature of WSNs make it quite difficult to recharge node batteries. Therefore, energy efficiency is a major design goal \cite{younis2006node}. In view of striking balance, relay nodes (RN)s with less complicated circuitry but powerful transceivers are generally employed to foster the overall energy efficiency of network since their transmission and reception capabilities are better than sensor nodes \cite{Cheng:2008}. It is worth noting that RNs are expensive than sensor nodes because they have more powerful transceivers and stronger battery banks. This is why a direct consequence of minimizing usage of RNs yields substantial reduction in overall WSN deployment cost.\\*[.3pc]
\indent
In most applications, nodes are generally patterned in three-dimensional fashion and lifetime enhancement with cost and connectivity constraints becomes an arduous task. This is because network's traffic is in $3D$, that is, in x-y-z planar surfaces \cite{Turjman:2012,Al-Turjman:2010,Son:2006}. Previous studies have shown that deployment in large $3D$ space is non-trivial. It has equally been shown that the combinatorial search space of this deployment problem is huge and each position translates to a different connectivity level. Therefore, the choice of optimization technique is paramount to ensure convergence \cite{Bari:2007}. In general, the computational efficiency is often enhanced using heuristics \cite{Turjman:2012}. These heuristics, especially the biologically inspired ones, are ubiquitously used to proffer solutions to a wide variety of engineering problems. Recently, promising evolutionary algorithms such as Artificial Bee Colony (ABC) \cite{Karaboga:2007}, Genetic Algorithm (GA) \cite{gupta2016genetic}, Gravitational Search Algorithm (GSA) \cite{rashedi_gsa:_2009}, Particle Swarm Optimization (PSO) \cite{magan2016optimal}, and Differential Evolution (DE) \cite{storn_differential_1997} have efficiently solved complex contemporary problems with high computational cost.\\*[.3pc]
\indent
RN deployment in WSN applications is broadly categorized into grid and non-grid-based deployment. In grid-based deployment, precise localization and data measurements are achievable because nodes are deployed on predefined grid vertices. In non-grid-based approaches, nodes are deployed in a multi-dimensional ad hoc manner. In practical 3-D settings, grid-based approaches are generally preferred to non-grid because the infinite search space in non-grid-based approach is reduced into a reasonably finite one. It has been shown that one way to tackle connectivity issues in WSNs is by deploying redundant nodes to repair connectivity when network partitioning occurs \cite{Cerpa:2004}. Redundant nodes were turned "on" only when the network connectivity was below a predefined threshold. Node's mobility has also been used to improve connectivity of a disconnected network \cite{Marta:2009}. Deployment area was modeled as a grid with cells of equal size and few number of RNs are carefully populated to connect the disjointed WSN segments. Network optimization is performed in such a way that RNs are deployed in least number of cells till connection is restored in all the segments \cite{Lee:2006,Lee:2010}.\\*[.3pc]
\indent
In \cite{patel2005energy,capone2010deploying}, integer linear programming was utilized to determine the optimal positions of sensor node, relay nodes, and base station (BS) that guarantee desired coverage, connectivity, bandwidth, and robustness. Since these deployments rely on the linearity property that is of importance in linear programming, it will be difficult to capture the nonlinear dynamics that can be introduced due to effect such as parameter variations especially in rugged terrains. Also, a two-tier two-phase architecture where sensor nodes are logically separated from backbone devices was proposed to repress the complexity of RN deployment \cite{Turjman:2012,Abbasi:2009}. In this two-tier strategy, sensor nodes are solely responsible for sensing and transmitting data to their corresponding cluster head (CH) or the closest RN. Hence, sensor nodes occupy the first layer of the architecture. In this design, sensor nodes are made to sleep when idle to minimize their energy consumption. In the first deployment phase, backbone was connected by deploying first-phase relay nodes (FPRNs) on candidate vertices in a 3-D fashion using Minimum Spanning Tree (MST). The purpose of deploying these FPRNs is to initiate connections among CHs and BS whose positions are fixed. In the second deployment phase of Shortest-Path-3D grid deployment (SP3D), desired connectivity was achieved by randomly deploying more RNs near the backbone devices.\\*[.3pc]
\indent
Optimized 3-D deployment with Lifetime Constraints (O3DwLC) \cite{Turjman:2012} is an improvement on SP3D and was developed to enhance network connectivity. The main difference between between SP3D and O3DwLC is evident in the way RNs are deployed in the second phase. In second phase of O3DwLC strategy, a semi-positive definite optimization algorithm was used to efficiently enhance the network connectivity subject to cost and lifetime constraints. It is worth mentioning that the proposed approach in this paper differs from those in SP3D and O3DwLC in the sense that it focuses on using minimum number of relay nodes to maximize the useful WSN lifetime with meta-heuristics approach. A two-tier constrained RN placement was proposed to improve the energy efficiency of a single-cover problem scenario, however, the solution utilizes more relay nodes than necessary \cite{10.1109/TMC.2011.126}. In addition, if the candidate positions are known a priori, a minimum number of relay nodes are capable of harvesting energy that can be utilized to improve both the connectivity and survivability of the network \cite{6583152}. Even though the idea of using partition detection/recovery algorithm and relay node mobility to detect and renew the connectivity of partitioned network sounds interesting, however, this type of solution involves mobile robots and other electronic circuitry, which could render it cost ineffective to implement especially in rugged terrains \cite{10.1109/TC.2009.120}.\\*[.3pc]
\indent
DE, GSA, and ABC have been shown as promising evolutionary tools for finding global optima for a variety of optimization problems with large search space. Many research disciplines, such as cloud manufacturing \cite{zhou2016hybrid}, bioinformatics \cite{karaboga2016discrete,ogundijo2017bayesian,ogundijo2016reverse}, electrical dispatch \cite{neto2017solving}, and big data \cite{sabar2017heterogeneous,han2017novel} have successfully adopted these evolutionary techniques for solving different non-trivial search problems. In general, optimization techniques converge to optimal solution in search space with predefined size, where the size is defined by a number of optimized parameters satisfying some constraints. Evolutionary algorithms are heuristics
and it is worth noting that there is no one-size-fits-all evolutionary technique for all optimization problems \cite{hashim2015fuzzy}. Some heuristics converge faster to better solutions than others in some problems and slower in other problems. Therefore, the three benchmark algorithms in this work belong to different search families. DE relies on genetic mutation and crossover \cite{storn_differential_1997}, ABC uses the roles of three main groups within a bee colony \cite{Karaboga:2007} and GSA is based on gravitational laws and mass interactions between particles \cite{rashedi_gsa:_2009}.\\
\indent
These optimization algorithms have been used in different contexts to optimize various capacities of WSNs. Coverage and connectivity have both been maximized using territorial predator scent marking algorithm; the algorithm also guarantees least overall energy consumption \cite{Abidin:2014}. ABC and Particle Swarm Optimization have also been utilized to solve sensor nodes placement and lifetime maximization problem \cite{Mini:2014}. In addition to all the aforementioned heuristic based applications, ant colony with greedy migration has also been utilized to deploy sensor nodes with a view to maximizing coverage and minimizing network deployment cost. A multi-objective approach has also been utilized to solve RN placement problem by optimizing three conflicting objectives - network reliability, cost, and average sensitivity area \cite{lanza2015assuming}. Unlike \cite{Udgata:2011,Mini:2014,Abidin:2014} that mainly focused on finding the optimal sensor node positions that enhance the network lifetime, heuristic-based RN deployment here proposed focuses on improving the network lifetime by deploying fewest number of RNs while satisfying connectivity constraint . Other closely related work to ours is that in \cite{magan2016optimal} that focused on optimal relay node placement in multi-hop scenario. In their approach, a two-step procedure and optional third step were utilized to optimize the network connectivity and throughput using PSO algorithm. The key difference compared to our method is that our method aims at optimizing network lifetime with lower bound constraints on network connectivity and upper bound constraint on cost of deployment.\\*[.3pc]
\indent
Lifetime maximization with cost and connectivity constraint problem introduced in \cite{Hashim:2015}, formulated using a two-phase two-tier approach. The approximate solution of the problem addressed has been shown to be NP-hard \cite{Bari:2007,Efrat:2008}. The complexity was extenuated using a two-phase two-tier strategy. The connectivity of network backbone was achieved by deploying least number of first-phase RNs for cost effectiveness \cite{Hashim:2015}. Evolutionary-based heuristic was utilized in the second phase to search for near-optimal positions of RNs. Optimization of network parameters was implemented to satisfy minimum application specific connectivity level constraint and guarantee minimum objective function. The placement problem was formulated in such a way that network cost and connectivity were simultaneously constrained in the desired range. ABC was employed as the evolutionary heuristic to search the optimal positions. In this work, the main contributions are: (i) the objective function in \cite{Hashim:2015} has been reformulated using additional penalty term for optimal performance, (ii) two other evolutionary techniques (DE and GSA) have been used in the second phase to efficiently deploy the RNs, (iii) the effectiveness of the proposed algorithms is compared and contrasted with \cite{Hashim:2015} on the basis of network lifetime enhancement and speed of convergence, and lastly (iv) comprehensive experiments have been carried out to show the efficacy (faster convergence and better optimal solution) of using DE as opposed to ABC presented in \cite{Hashim:2015} to deploy backbone devices in WSNs. \\
\indent
The rest of this paper is organized as follows: Section 2 presents the formulation of the problem statement . Section 3 describes the proposed deployment scheme and its mapping into an optimization problem. Section 4 describes the software implementation and results. Section 5 concludes the main findings and observations.
\section{Network Model and Placement Problem Formulation}
This section delineates the proposed network model. Hierarchical network architecture is assumed to address node heterogeneity problem. Graph topology is considered for easy network abstraction and effective computation of inter-node distances.\\*[-2.3pc]
\subsection{Assumptions}
\begin{figure}[H]
  \centering
  \includegraphics[width=8cm]{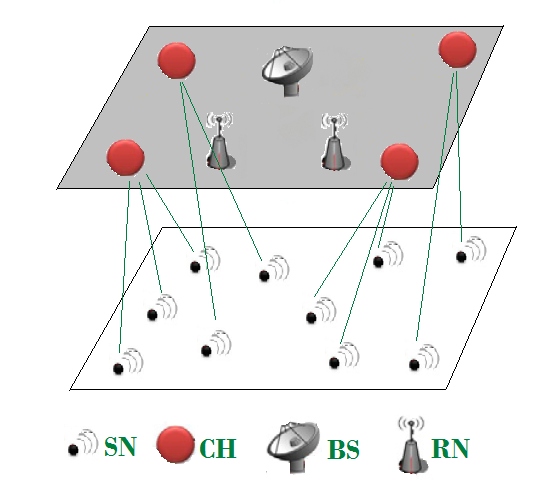}
  \caption{Two-tiered hierarchical scheme.}\label{twolayer}
\end{figure}

The network considered in this work mainly comprises of sensor nodes, CHs, RNs, and BS. Network heterogeneity is addressed using a two-tier hierarchical architecture shown in Fig.~\eqref{twolayer}. In the adopted design, sensor nodes occupy lower layer and they primarily collect data of interest and relay information to upper layer devices (RN or CH). It is assumed in this architecture that sensor nodes are only used for data sensing and transmission over short distances and that nodes sleep when idle. The CHs, RNs, and BS constitute the upper layer building blocks. Upper layer devices have stronger transceiver circuitry to transmit and receive data over longer distances. These devices also relay information from sensor nodes periodically to the base station.\\*[.3pc]
\indent
It is also assumed in this work that sensor nodes have enough energy to perform their sensing duties. This assumption gives the flexibility to focus on the upper layer devices. It has also been established that data sensing and processing requires lesser energy consumption than data transmission and reception \cite{Santamaria:2010,Olariu:2006}. Thus, energy consumed during data transmission and reception is only considered.
In addition, this two-tier approach decouples energy utilization of upper layer from lower layer as shown in Fig.~\eqref{twolayer}. Other assumptions in this work are as follows:
\begin{itemize}
  \item Nodes in the network are considered to be static.
 % \item Sensed data from nodes can either be transmitted to CHs or RNs and to another CHs or RNs and then to BS.
  \item Multi-hop communication is considered.
  \item The transmission and reception range of all CHs are the same.
  \item The transmission and reception range of all RNs are the same.
  \item The Media Access Control (MAC) protocol is ideal, that is, no packet collision and retransmission is experienced.\\*[-1pc]
\end{itemize}
\subsection{Energy Utilization Model}
The adopted two-tier architecture and the assumption that sensor nodes have enough resources to perform their tasks allows the isolation of lower layer energy utilization. This affords the flexibility to focus on the energy utilization of upper layer devices. Therefore, on the energy utilizations of RNs, CHs, and BS are considered. The energy consumption model proposed in \cite{Xu:2010,Rodrigues:2007} is also utilized in this work. Energy consumption of receiving device is:
\begin{equation} \label{MyEq1}
  \begin{split}
  J_{rx} &= L\beta,
  \end{split}
 \end{equation}
and that consumed by transmitting device is given by:
 \begin{equation} \label{MyEq2}
  \begin{split}
  J_{tx} &= L(\epsilon_1+\epsilon_2 d^{\gamma}),
  \end{split}
 \end{equation}
where $\beta$, $\epsilon_1$ and $\epsilon_2$ are transceiver hardware parameters \cite{Turjman:2012}, $L$ is the length of a packet, $d$ is the Euclidean distance between the transmitter and receiver, and $\gamma$ is the path loss exponent.
Remaining energy $E_{r}$, is given by:
 \begin{equation} \label{MyEq3}
  \begin{split}
  E_{r} &= E_i-TJ_{tx}-RJ_{rx}-\underline{A}J_a
  \end{split}
 \end{equation}
where $T$, $R$ and $\underline{A}$ are the arrival rates of transmitted, received and aggregated packets per round, respectively; $E_{i}$ is the initial energy of each node; $J_{tx}$, $J_{rx}$, and $J_a$ are energy per unit time consumed by each node for a single packet transmission, reception and aggregation, respectively.\\*[.3pc]
\indent
The energy mode in this work also incorporates the effect of packets transmitted by a RN to BS on behalf of other RNs. It is remarked that this effect depends on how close a RN is to BS. The closer a RN is to BS, the more the likelihood of relaying more data and the faster it depletes its energy. This subset of RNs is prone to causing prematurely network partitioning than other nodes thereby leading to early performance degradation. Incorporating this effect in the energy consumption model yields:
\begin{equation} \label{MyEq4}
  \begin{split}
  E_{p} &= kTJ_{tx}+ kRJ_{rx} + \underline{A}J_{a}
  \end{split}
\end{equation}
 where $k=$ packets received and relayed by a RN from other RNs.\\*[-.1pc]
By letting the initial rounds $I_R$ denote the total number of rounds the first-phase RNs can stay operational, overall network lifetime (in rounds) is derived as follows:
 \begin{equation} \label{MyEq12}
  \begin{split}
   I_R= \tfrac{B_1}{\sum_{p=1}^{FPRN}E_p}
  \end{split}
 \end{equation}
 where $B_1$ $=$ network energy after deploying FPRN
 \begin{equation} \label{MyEq13}
  \begin{split}
   Lifetime= T_R= \tfrac{\sum_{i=1}^{2}B_i}{\sum_{p=1}^{FPRN+SPRN}E_p}
  \end{split}
 \end{equation}
 where
 \begin{equation} \label{MyEq14}
  \begin{split}
   E_{p} = kTL(\epsilon_1+\epsilon_2 \mu_{w} ^{\gamma})+ kRL \beta + \underline{A}J_{a},
  \end{split}
 \end{equation}
 with $\mu_{w}$ denoting the average internode distance, and $B_2$ being the energy introduced by deploying SPRN. Then network lifetime (in rounds) is given as:
 \begin{equation} \label{MyEq15}
  \begin{split}
   Lifetime= T_R= I_R + \sum_{i=1}^{n_c}\alpha_i E_R
  \end{split}
 \end{equation}
where $E_R =$ extra rounds achievable due to addition of one second-phase RN to a candidate position; $n_c$ is the number of candidate positions a second-phase RN can be placed on the grid and $\alpha_i$ is 1 if second-phase RN is placed on vertex $i$ and 0 otherwise.
\subsection{Network Lifetime Definition}\label{defn}
Wireless sensor network lifetime is defined in this work as the maximum number of rounds achievable before network partitioning occurs, that is, the duration in rounds when most RNs and CHs have depleted their energy and are no longer able to relay packets to the next hop or sink. This lifetime definition can also be interpreted as the duration from deployment to instance when the algebraic network connectivity and total energy fall below some partition thresholds. The assumption that network connectivity and/or lifetime directly depend(s) on the activity of CHs/RNs originates from the two-layer architecture adopted. The connectivity of network backbone is highly dependent on the connectivity of RNs. In practice, energy partition threshold can vary from application to application and from network to network. These variations are due to factors such as power rating of the transceivers, node's density and position of RNs within a network. It is also noted that the definition adopted in this paper does not take cognizance of such factors. In order to avoid threshold variations within the network, partition threshold is set based on the total energy in network backbone and not on the energy of individual RN.
\subsection{Problem Statement}
By predefining the positions of sensor nodes, CHs, BS, and first-phase RNs, the aim is to find the positions of pre-determined number of second-phase RNs (SPRNs) on the grid that enhance the overall network lifetime. That is, if given the grid locations of sensor nodes, CHs, first-phase RNs, and BS, the objective is to find near-optimal positions of predetermined number of SPRNs on 3-D grid that maximize the network lifetime while satisfying connectivity constraints. The network cost is constrained by ensuring the number of SPRNs do not be exceed a predefined number. The second phase of the proposed deployment strategy is where RNs are strategically deployed using optimization heuristics. The positions of SPRNs are optimized using two algorithms that are known for searching global optimum. The objective in this phase is to minimize the average internode distances, $\mu_w\left(G\right)$ with constraints on the number of SPRNs that can be deployed and the algebraic network connectivity. This optimization problem is then formulated as:
\begin{equation} \label{MyEq9}
  \begin{split}
   \min_{\alpha} \left(\mu_{w}\left(L\left(\alpha\right)\right) + \beta \max \{0, \left(\lambda_{2,FPRN}-\lambda_{2,FPRN+SPRN}\right)\}\right), \\
   s.t \ \ \ \ \sum_{i=1}^{n_c} \alpha_i = N_{SPRN},\ \ \ {\rm where} \ \ \ \ \alpha_i\in [0,1],
  \end{split}
\end{equation}

 \begin{equation} \label{MyEq11}
  \begin{split}
   L\left(\alpha\right)=L_i+ \sum_{i=1}^{n_c}\alpha_i A_i A_i^T
  \end{split}
 \end{equation}
where $A_i$ is the incidence matrix generated by adding SPRNs, $L_i$ is the initial Laplacian generated using MST; $n_c$ is the number of candidate grid locations for SPRNs and $\alpha_i$ is 1 if SPRN is placed on vertex $i$ otherwise 0. The mathematical derivations of overall network lifetime is given in Eq. \eqref{MyEq4}-\eqref{MyEq15}. $\beta$ controls the importance imposed on connectivity constraint and was experimentally set to $1$. As a test case, $\lambda_{2,FPRN}$ was set to $0.5$.\\
\indent
It is worth mentioning that the penalty term added to the objective function in Eq. \eqref{MyEq9} enforces the connectivity constraint. As opposed to a hard connectivity constraint imposed in \cite{Hashim:2015}, a soft constraint is here enforced and it is remarked that soft constraint gives the optimizer flexibility necessary to find good local optimal in the objective function landscape.
\section{The Meta-heuristic Based Deployment Strategies}
\begin{figure}[h]
  \centering
  \includegraphics[width=8cm]{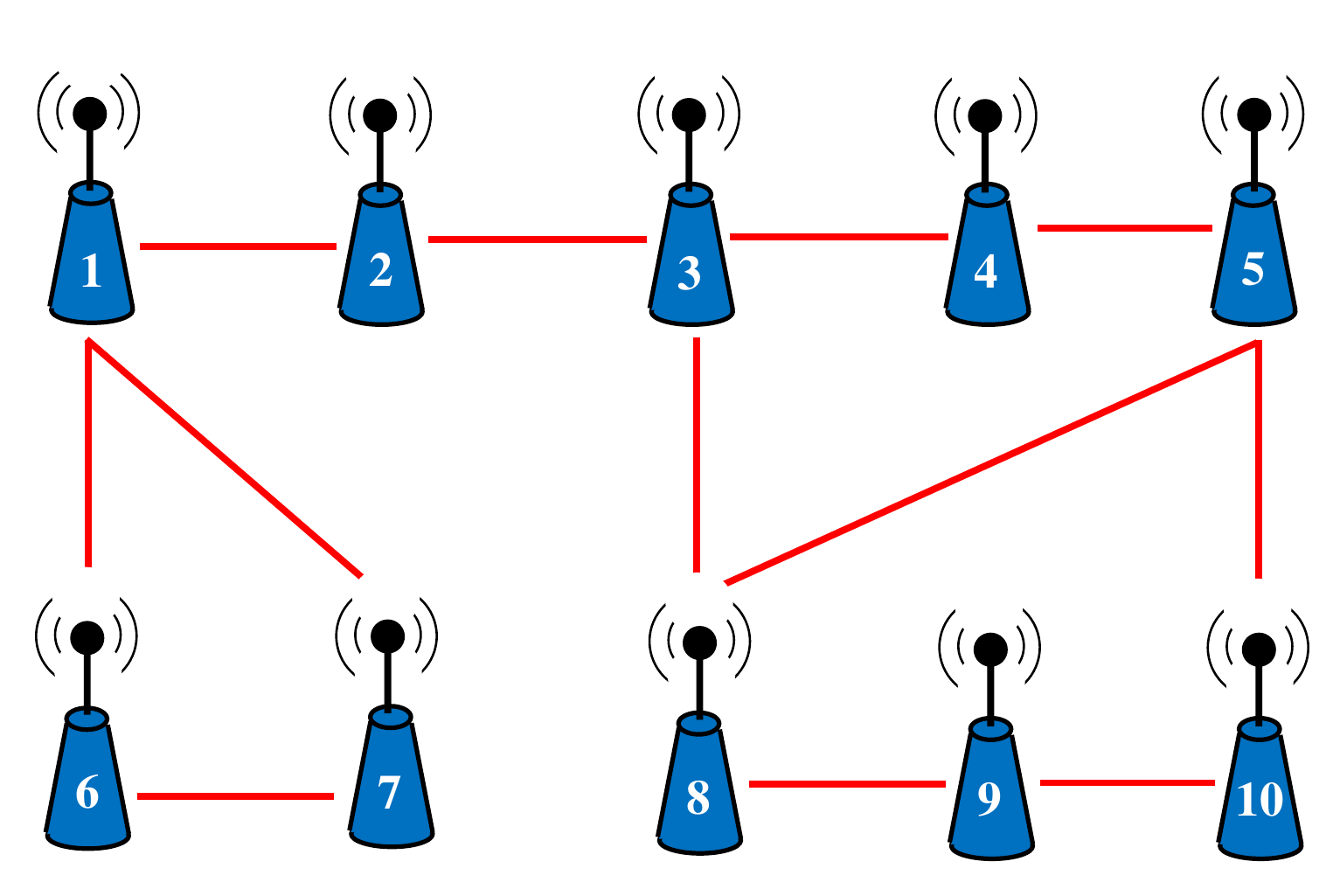}
  \caption{An example of a 10-node-12-link graph.}\label{laplacian}
\end{figure}

In this section, RN deployment strategy using evolutionary optimization tools is described in detail. The two optimizers are objectively compared and contrasted with ABC in terms of their effectiveness in finding optimal SPRN grid positions and in terms of convergence speed. It is worth noting that the NP-Hard properties of this deployment problem can be circumvented using a two-tier two-phase architecture, which is similar to what was adopted in \cite{Turjman:2012,Hashim:2015}. In the first phase of deployment, MST constructs the network backbone using first-phase RNs (FPRNs) and ensures that RNs, CH, and BS are connected to a desired connectivity level \cite{Turjman:2012}. The pseudocode of MST is given in Algorithm~1 as implemented in \cite{Turjman:2012}.\\*[.3pc]
\begin{algorithm}
\caption{MST: FPRN Deployment (Construct Backbone $B$)}
\label{alg:MST}
\begin{algorithmic}[1]
   \STATE  {\bf Initialize} :
   \STATE  {\ \ \ \ \ \ \ \ Initial number of CHs and BS to construct $B$};
   \STATE  {\bf Input} :
   \STATE  {\ \ \ \ \ \ \ \ A set IS of the CHs and BS nodes coordinates};
   \STATE  {\bf Output} :
   \STATE  {\ \ \ \ \ \ \ \ A set CC of the CHs, minimum RNs, and BS coordinates forming}
   \STATE  {\ \ \ \ \ \ \ \ the network Backbone};
   \STATE  {\bf begin}
   \STATE  {\ \ \ \ \ \ \ \ CC $=$ set of closest two nodes in IS;}
   \STATE  {\ \ \ \ \ \ \ \ CC $=$ CC $\cup$ minimum RNs needed to connect them on the 3-D grid};
   \STATE  {\ \ \ \ \ \ \ \ IS $=$ IS - CC};
   \STATE  {\ \ \ \ \ \ \ \ CC = CC [ minimum RNs needed to connect them on the 3-D grid};
   \STATE  {\ \ \ \ \ \ \ \ $N_d$ $=$ number of remaining IS nodes which are not in CC};
   \STATE  {$i = 0$};
   \STATE  {\bf for} each remaining node $n_i$ in IS.
   \STATE  {\ \ \ \ \ \ \ \ Calculate $M_i$: Coordinates of minimum number of RNs required}
   \STATE  {\ \ \ \ \ \ \ \ to connect $n_i$ with the closest node in CC}.
   \STATE  {\ \ \ \ \ \ \ \ $i = i + 1$};
   \STATE  {\bf end.}
   \STATE  $M = {Mi}$
   \STATE  {\bf while} $Nd > 0$ do :
   \STATE  {\ \ \ \ \ \ \ \ SM $=$ Smallest $M_i$};
   \STATE  {\ \ \ \ \ \ \ \ CC$=$ CC $\cup$ SM $\cup$ $n_i$};
   \STATE  {\ \ \ \ \ \ \ \ IS $=$ IS-$n_i$};
   \STATE  {\ \ \ \ \ \ \ \ M $=$ M - $M_i$};
   \STATE  {\ \ \ \ \ \ \ \ $N_d $=$ N_d - 1$};
   \STATE  {\bf end.}
\end{algorithmic}
\end{algorithm}
%\\*[.2pc]
%\indent
If network backbone $B$ is perceived as a connected graph ($G$) then from graph theory, Laplacian ($L$) can be constructed from $G$ \cite{Ghosh:2010,Bhardwaj:2001,Boyd:2006}. $L$ is symmetric, semi-positive definite, and element $\left(i,j\right)$ has a value of $-1$ if node $i$ can receive and send packets to node $j$. The principal diagonal elements L$\left(i,i\right)$ have positive integer numbers that denote the number of RNs connected to node $i$ \cite{Turjman:2012}. The ordering of eigenvalues of $L$ is given in Eq. \eqref{MyEq5} below:
\begin{equation} \label{MyEq5}
  \begin{split}
  0=\lambda_1 \leq \lambda_2 \leq \lambda_3 \leq ....\lambda_{n-1} \leq \lambda_n,\\
  \end{split}
 \end{equation}
where $\lambda_2$ is known as the Fielder value. Note that $\lambda_2 = 0$ when there is no connectivity in $G$. By implication, the algebraic connectivity of backbone $G$ is characterized by $\lambda_2$. It is remarked that connectivity level of the network can be tuned to a desired level by strategically augmenting the number of FPRNs using the MST. On the other hand, $\lambda_2 = 1$ denotes when the network is well connected, that is, all the nodes have direct connections to every other nodes. An example of a connected 10-node-12-link backbone (or graph) is given in Fig.~\ref{laplacian}. It is worthy of note that removing one node may partition the entire network. For instance, node 3 in Fig.~\ref{laplacian} is a typical illustration of a node whose removal could partition the network backbone. The Laplacian matrix ($L$) generated from Fig.~\ref{laplacian} is given in \eqref{MyEq5a} with specific $\lambda_2$. The Fielder value ($\lambda_2$) of the matrix ($L$) denotes the link count needed for network partitioning \cite{Turjman:2012}.

\begin{equation} \label{MyEq5a}
L_{10}= {\left(\begin{smallmatrix} 3 & -1 & 0 &0&0&-1&-1&0&0&0\\ -1 & 2 & -1 &0&0&0&0&0&0&0\\ 0 & -1 & 3 &-1&0&0&0&-1&0&0\\0 & 0 & -1 &2&-1&0&0&0&0&0\\0 & 0 & 0 &-1&3&0&0&-1&0&-1\\-1 & 0 & 0 &0&0&2&-1&0&0&0\\-1 & 0 & 0 &0&0&-1&2&0&0&0\\0 & 0 & -1 &0&-1&0&0&3&-1&0\\0 & 0 & 0 &0&0&0&0&-1&2&-1\\0 & 0 & 0 &0&-1&0&0&0&-1&2 \end{smallmatrix}\right)}
%\]\\
\end{equation}
The following definitions in \cite{Merris:1990} are adopted in this work:
\begin{enumerate}\label{Enume1}
                  \item Let $G$ be a graph with vertex set $V\left(G\right)$ and edge set $E\left(G\right)$. The distance $d_G\left(u,v\right)$ between two vertices $u,v \in V \left(G\right)$ is the minimum number of edges on a path in $G$ between $u$ and $v$.
                  \item $d_G\left(v\right)$ is defined as the distance of a vertex $v$ in $G$ and it is given as the sum of all distances between vertex $v$ and all other vertices of $G$.
                  \item The Wiener index $W\left(G\right)$ is a property of graph $G$ and is related to $d_G\left(v\right)$ through :
                \end{enumerate}
\begin{equation} \label{MyEq6}
 \begin{split}
   W\left(G\right) &= \tfrac{1}{2}\sum_{v\in V\left(G\right)}d_G\left(v\right)=\tfrac{1}{2}\sum_{\left(u,v\right)\in V\left(G\right)}d_G\left(u,v\right)\\
  \end{split}
\end{equation}
where $\tfrac{1}{2}$ compensates for the fact that distance between $u$ and $v$ are summed twice since $G$ is undirected. On averaging the distance between every pair of vertices of $G$, we obtain the average distance $\mu\left(G\right)$:
 \begin{equation} \label{MyEq7}
 \begin{split}
  \mu\left(G\right) &= \tfrac{W\left(G\right)}{\tfrac{n}{2}}=\tfrac{W\left(G\right)}{\left(
                                              \begin{array}{c}
                                                \lvert V\left(G\right)\rvert\ \\
                                                2 \\
                                              \end{array}
                                            \right)
  }\\
 \end{split}
 \end{equation}
Eq. \eqref{MyEq7} is modified to Eq. \eqref{MyEq7b} in order to incorporate any unmodeled variation in the average distance. Unmodeled variation is added to the cost function as a constant penalty term. This gives the proposed solution a level of robustness - a feature that makes it fit into many real life scenarios.
\begin{equation} \label{MyEq7b}
 \begin{split}
  \mu_{w} \left(G\right) &= \tfrac{W\left(G\right)}{\tfrac{n}{2}}=\tfrac{W\left(G\right)}{\left(
                                              \begin{array}{c}
                                                \lvert V\left(G\right)\rvert\ \\
                                                2 \\
                                              \end{array}
                                            \right)
  }+\Delta\mu\left(G\right)\\
 \end{split}
 \end{equation}
$\Delta\mu\left(G\right)$ is added to account for unpredictable occurrences such as slight displacement of nodes by surrounding air. It is however important to establish its upper bound for convergence.\\
By ordering the eigenvalues of Laplacian as $\lambda_1 \leq \lambda_2 \leq \lambda_3 \leq ....\lambda_{n-1} \leq \lambda_n$, the computation of Wiener index is done using:
\begin{equation} \label{MyEq8}
  \begin{split}
   W\left(G\right) &= n \sum_{2}^n\frac{1}{\lambda_i}\\
  \end{split}
 \end{equation}

\subsection{Optimization Algorithms}
Each optimization algorithm aims at finding the minimum objective function as defined in Eq. \eqref{MyEq7b}, which corresponds to optimal positions of SPRNs. The following subsections briefly describe the two optimization techniques implemented in the proposed algorithm.
\subsection{Differential Evolution}
\begin{figure}[!h]
  \centering
  \includegraphics[scale=0.5]{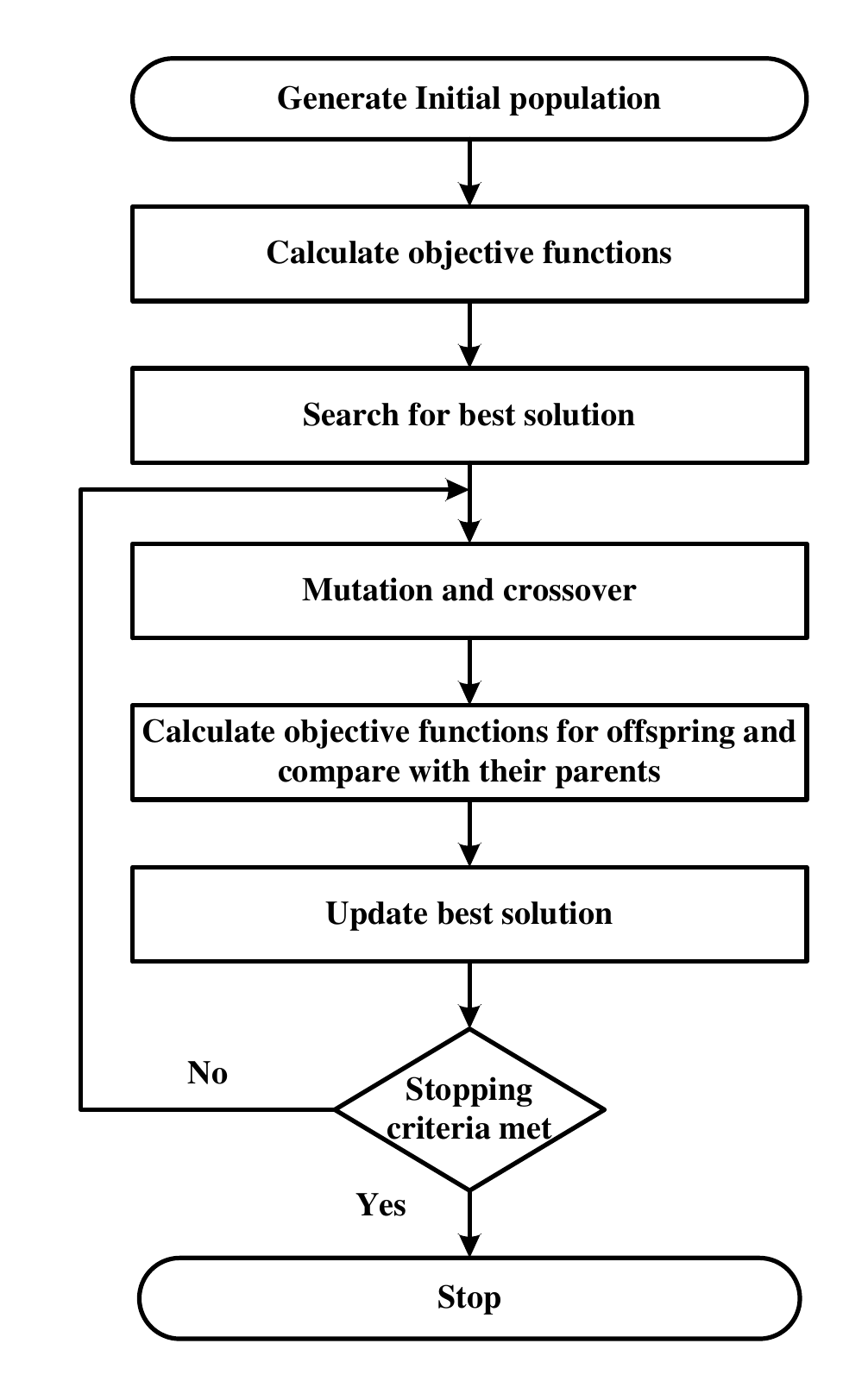}
  \caption{DE computational flowchart.}\label{Fig_4_DE}
\end{figure}

Differential evolution is a technique that was initially developed and tested on "Chebychev Polynomial fitting problem", which later metamorphosed into an impressive optimization technique \cite{storn_differential_1997}. In addition to retaining a better population and solution, DE uses the same structure as Genetic Algorithm such as crossover and mutation. In the mutation and crossover search process, important relations are used. Mutation requires assigning an arbitrary constant between 0 and 1 called the mutation probability (MP), and mutation relation is calculated only if MP is greater than a random number between 0 and 1 as follows:
\begin{equation} \label{MyEq100}
    v_i\left(t+1\right) = x_i\left(t\right) + \mathcal{F}\left(x_{best}\left(t\right) - x_i\left(t\right)\right)  + \mathcal{F}\left(x_{r_1}\left(t\right)- x_{r_2}\left(t\right)\right)
 \end{equation}
The crossover is evaluated using a relation where crossover probability (CP) is arbitrarily set to a value between 0 and 1, and then compared to a random number between 0 and 1. The crossover step is executed only if CP is greater than the random number. The crossover equation is then calculated using the relation:
\begin{equation} \label{MyEq101}
  \begin{split}
    x_i\left(t+1\right) = v_i\left(t+1\right)
  \end{split}
 \end{equation}
where $i = 1, 2, \ldots, N$ and $i$ is iteration number for every solution in the generation;  $\mathcal{F}>0$ is a constant factor; $x_i\left(t\right)$ denotes a solution at iteration $i$ in the generation; $v_i\left(t+1\right)$ is a mutant vector given in Eq. \eqref{MyEq100}; $x_{r_1}\left(t\right)$ and $x_{r_2}$ are solution vectors selected randomly from current generation; $x_{best}\left(t\right)$ is the best solution. A more detailed description of DE can be looked up in \cite{storn_differential_1997}. Summary of the algorithm is shown in the flowchart Fig. \ref{Fig_4_DE}.

\subsection{Gravitational Search Algorithm}
\begin{figure}[!h]
  \centering
  \includegraphics[scale=0.5]{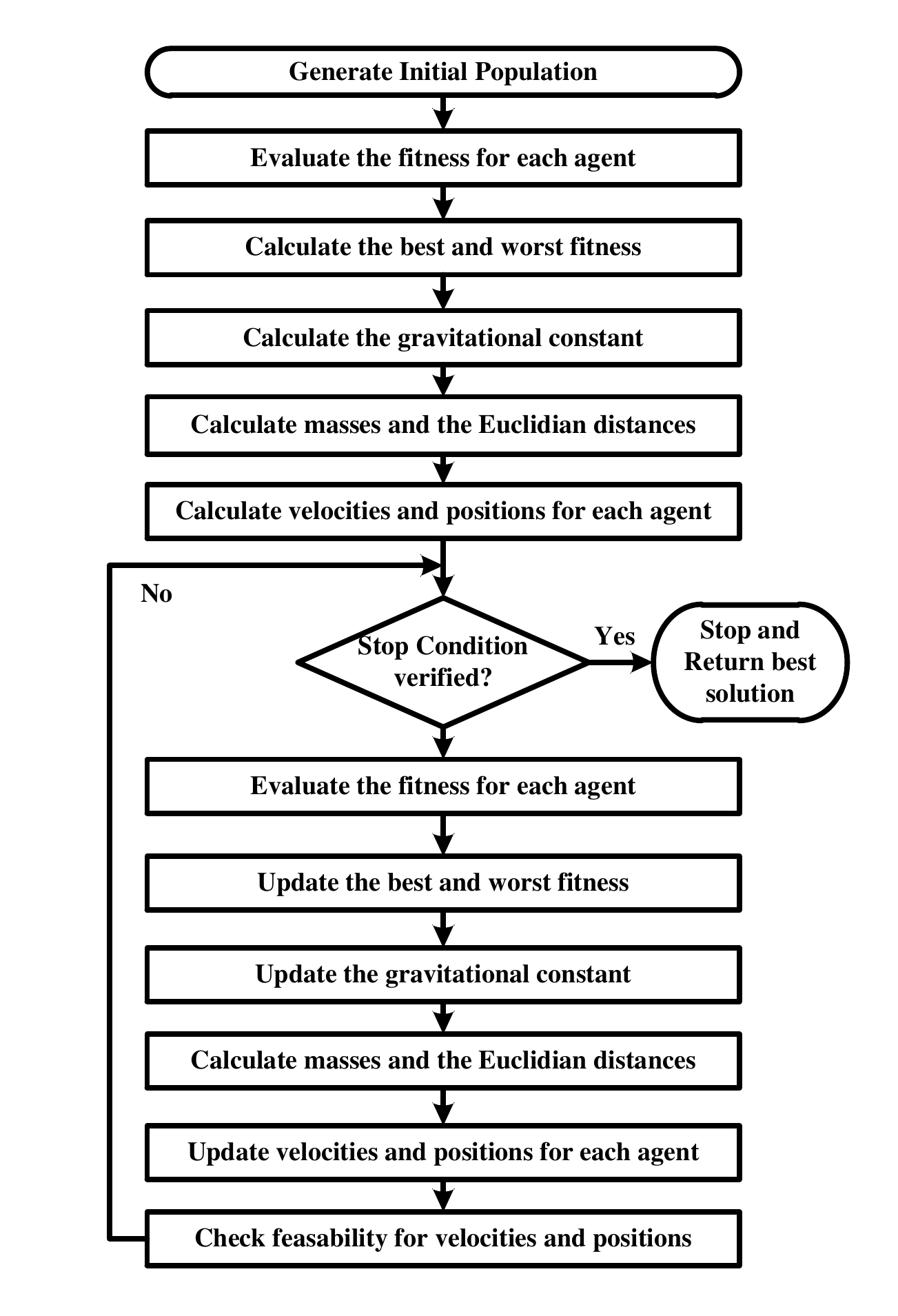}
  \caption{GSA computational flowchart.}\label{Fig_5_GSA}
\end{figure}

Gravitational search algorithm (GSA) is an important meta heuristic optimization algorithm that is based on Newton's laws of gravitation and it was initially proposed in 2009 \cite{rashedi_gsa:_2009}. The algorithm follows the gravitational law which stipulates that: "for any two objects, every object is attracted to the other object by a force which is directly proportional to their mass and inversely proportional to their square distance". According to the principle of gravity, the force between any two nodes can be evaluated by
\begin{equation} \label{MyEq201}
  \begin{split}
    F = G_c\frac{M_1M_2}{R^2}
  \end{split}
 \end{equation}
 where $M_1$ and $M_2$ are masses of the particles, $G_c$ is the gravitational constant, $R$ is the Euclidean distance between the two particles and $F$ is the gravitational force. It should be noted that $G_c$ is defined in the GSA algorithm by
\begin{equation} \label{MyEq202}
  \begin{split}
    G_c\left(t\right) = G_c\left(t_{0}\right){\rm exp}\left(-\frac{\alpha t}{T}\right)
  \end{split}
 \end{equation}
 such that $G_c\left(t\right)$ is the gravitational constant at any time $t$, $G_c\left(t_{0}\right)$ is the initial value of gravitational constant at time $t_0$, $\alpha$ is a positive constant and $T$ is final time, which can be assigned based on number of iterations. The particle acceleration is given by
\begin{equation} \label{MyEq203}
  \begin{split}
    a_{i}^{d}\left(t\right) = \frac{F_{i}^{d}\left(t\right)}{M_{i}\left(t\right)}
  \end{split}
 \end{equation}
 where $i = 1,\ldots,N$ and $d = 1,\ldots,D$ such that $N$ is number of particles and $D$ is number of optimized parameters within the particle. $F_{i}^{d}$ is force of particle $x_{i}^{d}$ and ${M_{i}}$ is the mass of particle $i$. Next, positions and velocities of each parameter can be defined according to the following relations
 \begin{equation} \label{MyEq204}
   \begin{split}
     v_{i}^{d}\left(t+1\right) = rand_iv_{i}^{d}\left(t\right) + a_{i}^{d}\left(t\right)
   \end{split}
  \end{equation}
\begin{equation} \label{MyEq205}
  \begin{split}
    x_{i}^{d}\left(t+1\right) =  x_{i}^{d}\left(t\right) + v_{i}^{d}\left(t+1\right)
  \end{split}
 \end{equation}
  where $x_{i}^{d}\left(t+1\right)$ and  $v_{i}^{d}\left(t+1\right)$ are the position and velocity of parameter $d$ in particle $i$ at time $t+1$ or at the next iteration, respectively. Also, $rand_i$ is a random number between 0 and 1. A detailed explanation of the GSA is given in \cite{rashedi_gsa:_2009} and the flowchart in Fig. \ref{Fig_5_GSA} gives the synopsis of the algorithm \cite{rashedi_gsa:_2009,Hash:2015}.
\subsection{Implementation of the Optimization algorithms }
The proposed deployment is formulated using optimization algorithms to update the initial Laplacian ($L_i$) obtained from MST algorithm. Using the observation that the algebraic network connectivity is the second least eigenvalue ($\lambda_2$) of $L$, the connectivity can be constrained within a desired bound during optimization using a penalty function. It is worth noting that $L$ is symmetric and the magnitude of the off-diagonal row or column elements sum up to their corresponding diagonal elements. By considering the probabilistic communication model, network connectivity is the likelihood that two nodes separated by distance $d$ can establish communication is:
\begin{equation} \label{MyEq15c}
  \begin{split}
 P_c = \varpi e^{-\mu d^\gamma}
  \end{split}
\end{equation}
where $d$ is the distance of separation between the transmitting and receiving nodes, $\gamma$ is the path loss exponent which is empirically determined, $\mu$ is a random variable that describes the attenuation encountered by signals upon deployment, and $\varpi$ is a constant estimated from the heights of transmitter, receiver, and monitored site. It can be deduced from Eq. \eqref{MyEq15c} that connectivity depends not only on separating distance between nodes but also on many other factors such as shadowing (due to obstacles) and multipath effects (due to terrain)  \cite{Turjman:2012}.\\
\indent
During the second phase deployment, our algorithm deploys SPRNs on candidate positions based on where the objective function is minimum and then updates the matrix $L$ accordingly as depicted in Eq. \eqref{MyEq11}. The objective function of the optimization is given in Eq. \eqref{MyEq8}. The aim of the optimizer is to  minimize the objective function subject to constraints on cost and $\lambda_2$. Both DE and GSA are separately used as tools for optimizing the network parameters. The algorithms optimize the off-diagonal elements of $L$ by terminating some connections with value 0 and/or reconnecting other with value -1. The algorithm ensures that the connectivity constraint is satisfied before a position is accepted as a feasible solution. The updated Laplacian matrix is used to evaluate the average distance among the RNs/CHs using Eq. \eqref{MyEq7}. The lifetime is then evaluated using the set of equations in the Eq. \eqref{MyEq4}-\eqref{MyEq15}.\\*[.4pc]
\indent
It must be remarked that one essential factor in the energy utilization model is the distance between the transmitting and receiving nodes as given in Eq.\eqref{MyEq14}. Since $\left(\mu_w\left(G\right)\right)$ is proportional to $W\left(G\right)$, the optimizers evaluate the fitness using $W\left(G\right)$. The size of $L$ is determined by the total number of backbone devices, that is, total number of RNs, CHs, and BS. The number of SPRNs is incremented and are placed on candidates vertices and the Laplacian matrix is updated. The major challenge of the optimizers is choosing best set of locations that result in least average inter-node distance. SPRNs are randomly placed on a subset of candidate vertices and the algorithm checks if the fielder value of the new $L$ is within the specified bounds before accepting or discarding a solution. This process continues until the solution converges. The best deployment with the smallest internode distances is returned by the algorithm. The nodes that are directly connected to the SPRNs will be visible using the most recently updated Laplacian matrix, $L$. The synopsis of the second-phase deployment is given in Algorithm \ref{alg:eldwcc}.
\begin{algorithm}
\caption{: Second-phase Deployment.}
\label{alg:eldwcc}
\begin{algorithmic}[1]
   \STATE  {\bf Input}:
       \STATE  {Locations of CHs, FPRNs and BS nodes.};
   \STATE  {\bf Output}:
   \STATE { Locations $X_P$ of the SPRNs that maximize lifetime of B with connectivity and cost constraints.}
   \STATE  {\bf begin}
   \STATE  $L_i$ $=$ Initial Laplacian from $B$;
   \STATE   $I_R$ $=$ Lifetime in the first-phase
   \STATE  $A_i$ $=$ adjacency matrix matching vertex i on the grid
   \STATE  Evaluate $\mu_w\left(G\right)$;
   \STATE  $E_R$ $=$ number of rounds added by deploying SPRN on vertex i;
   \STATE  $WC$ $=$ $X_P$ from either DE or GSA
%\EndFunction
\end{algorithmic}
\end{algorithm}
\section{Simulation Results}
Initial experimental setup in MATLAB consists of nine cluster heads and one base station on predetermined grid locations. Nine initial CHs are deployed very close to their respective cluster members since sensor nodes have limited energy and low capabilities to transmit over far distances. For this reason, sensor nodes are only required to gather information of interest from their immediate environment. In the proposed deployment, the number of RNs is varied from 10 up to 50 for both optimization techniques. The normalized average distance deviation $\left(\Delta\mu\left(G\right)\right)$ is set to 0.1 in all the experiments. For the purpose of comparing these techniques, number of rounds is used as figure of merit for lifetime enhancement. .\\*[.2pc]
\indent
It is remarked that the lifetime derived in Eq. \eqref{MyEq12}-\eqref{MyEq15} is the number (in rounds) the network stays functional without partitioning. However, in reality the rounds depends on various ingredients such as network area, data generation rate of the sensor nodes, and amount of traffic handled by RNs virtue of their locations relative to the sink. The results presented are normalized to give a clearer and more distinct notion of energy efficiency achievable using these evolutionary techniques. \\*[.4pc]
\indent
In all the experiments, the maximum number of generations is set to 200. The population depends solely on the network size, that is, the size of the population grows linearly with network size. For each optimizer and network size pair, experiment was performed eight times to mitigate the effect of random initialization. It must be noted that network size here refers to number of backbone devices and optimization parameters are the off-diagonal elements of the Laplacian. The synopsis of number of parameters and iterations as the network size grows is shown in Table~(\ref{table:result1}). Table~(\ref{table:resultcomp}) gives the initial settings of each of the algorithms. For DE, $MP$ is mutation probability, $CP$ is crossover probability, and $F$ is a factor between 0 and 1. For GSA, $\alpha$ is a positive constant that controls the gravitational speed, $\epsilon$ is a small positive constant, $\lambda$ is the division speed, $G_0$ is the gravitational constant, and $K_{best}$ is a set of best solutions. The results of the experiments are presented in Tables (\ref{table:result2a}) and (\ref{table:result2b}).\\*[.4pc]
\indent
\begin{table}[t] \label{Table1}
 \setlength{\tabcolsep}{1pt}
\caption{Network parameters for simulations} % title of Table
\centering % used for centering table
\small
\begin{tabular}{|c| c| c| c|} % centered columns (3 columns)
\hline\hline %inserts double horizontal lines
Parameters & Value & Parameters & Value\\ [0.5ex]
%heading
\hline\hline %inserts double horizontal lines
$n_c$ & $110$ & $L$ & 512 bits \\[1ex] % inserting body of the table
$J_a$ & $50 \times 10^{-7} J$ & $E_i$ & 15.4 J\\[1ex]
$\epsilon_1$ & $50 \times 10^{-9} J/bit$  & $T$ & 100(p/r) \\[1ex]
$\epsilon_2$ & $10 \times 10^{-12} J/bit/m^2$ & $R$ & 100\\[1ex]
$\gamma$ & $4.8$ & $A$ & 10\\[1ex]
$\beta$ & $50\times 10^{-9} J/bits$ & $r$ & 100m\\[1ex]
\hline\hline %inserts double horizontal lines
\end{tabular}
\label{table:nomen} % is used to refer this table in the text
\end{table}

%\begin{table*}[!t]
\begin{table*}[t]
 \setlength{\tabcolsep}{1pt}
\caption{Data for Optimization} % title of Table
\centering % used for centering table
\small
\begin{tabular}{|c| c| c| c|} % centered columns (3 columns)
\hline\hline %inserts double horizontal lines
Network Size (N) & No of Optimized Parameters& Size of Population& Iterations\\ [0.5ex]
%heading
\hline\hline %inserts double horizontal lines
20 & 190 & 400 & 200 \\[1ex] % inserting body of the table			
30 & 435 & 600 & 200\\[1ex]			
40 & 780 & 800 & 200 \\[1ex]			
50 & 1225 & 1000 & 200\\[1ex] 			
60 & 1770 & 1200 & 200\\[1ex]			
\hline\hline %inserts double horizontal lines
\end{tabular}
\label{table:result1} % is used to refer this table in the text
\end{table*}

\begin{table*}[!t]
 \setlength{\tabcolsep}{1pt}
\caption{Settings of optimization techniques} % title of Table
\centering % used for centering table
\small
\begin{tabular}{|c| c| c| c| c|c| c|} % centered columns (3 columns)

%heading
\hline\hline %inserts double horizontal lines        	   	   				
\multirow{2}{*}{DE} & Parameter &  \multicolumn{2}{|c}{$MP$}   & \multicolumn{2}{|c|}{$CP$}  &  $\mathcal{F}$   \\[1ex] % inserting body of the table			
\cline{2-7}
& Setting &  \multicolumn{2}{|c}{0.9} & \multicolumn{2}{|c|}{0.9}& 0.5	\\[1ex]   	   	   	   	   	   	   				 \cline{1-7}
\multirow{2}{*}{GSA}& Parameter & $\alpha$    &   $\lambda$    &   $\epsilon$     &   $ G_0$     &  $K_{best}$    \\[1ex]	  	  	  	  							 \cline{2-7}
& Setting &  7 &	6	 & 0.00001 &	1000	 & 4  \\[1ex]																	
\hline\hline %inserts double horizontal lines
\end{tabular}
\label{table:resultcomp} % is used to refer this table in the text
\end{table*}

\begin{table*}[!t]
 \setlength{\tabcolsep}{1pt}
\caption{Performance for network size $N=20$,$30$, and $40$} % title of Table
\centering % used for centering table
\small
\begin{tabular}{|c|c|c|c|c|c|c|c|} % centered columns (3 columns)
\hline %inserts double horizontal lines
\multicolumn{2}{|c|}{} & \multicolumn{2}{c|}{$N=20$} & \multicolumn{2}{c|}{$N=30$} & \multicolumn{2}{c|}{$N=40$}\\ [0.5ex]
\cline{3-8}
\multicolumn{2}{|c|}{} &  Mean $\pm$ STD & \textit{p}-value &  Mean $\pm$ STD & \textit{p}-value &  Mean $\pm$ STD & \textit{p}-value\\ [0.5ex]
%heading
\hline\hline        	   	   				
\multirow{5}{*}{ABC} & $W\left(G\right)$ &  20.8184 $\pm$ 0.6770 & 0.0703 & 47.3064 $\pm$ 1.3334 & \textless 0.0001 &  68.1976$\pm$ 2.0619 &  0.0177\\[1ex] % inserting body of the table			
\cline{2-8}
& $\mu\left(G\right)$ &  7.8046 $\pm$ 0.5682 & 0.0703 & 7.3611 $\pm$0.4888 &	\textless 0.0001  &  3.8544$\pm$0.4216	& 0.0177\\[1ex]  \cline{2-8}
& $E_p$ &  0.4927 $\pm$ 0.0485 & 0.0778 &  0.5275$\pm$ 0.0371 & \textless 0.0001 &  0.7196$\pm$0.0171 &  0.0239\\[1ex]	 \cline{2-8}
& $T_R$ &  4.2074 $\pm$ 0.6504 & 0.0620 & 7.9317$\pm$0.9917 & \textless 0.0001 &   27.9048$\pm$2.9905 & 0.0130 \\[1ex]  \cline{2-8} 		
& $\lambda_2$ & 0.5959 $\pm$ 0.0052 & 0.0037 &  0.5699$\pm$0.0052 & 0.8458  &  0.5689$\pm$0.0063 & 0.1534\\ \cline{1-8}

\multirow{5}{*}{DE}& $W\left(G\right)$ &  20.0059 $\pm$ 0.5023 & \text{-} &  37.3375 $\pm$ 1.7241 &  \text{-} &  63.9238$\pm$3.2922 &  \text{-}\\[1ex] % inserting body of the table			
\cline{2-8}
& $\mu\left(G\right)$ &  7.1226 $\pm$ 0.4216 & \text{-} &  3.7066$\pm$ 0.6320 &  \text{-} &   2.9807$\pm$0.6731	&  \text{-}	\\[1ex]  \cline{2-8}
& $E_p$ & 0.4927 $\pm$ 0.0308 & \text{-} &  0.7244$\pm$ 0.0227 &  \text{-} &  0.7491$\pm$0.0229 &  \text{-}\\[1ex]	  	  	  	  							 \cline{2-8}
& $T_R$ & 5.1173 $\pm$ 0.6080 & \text{-} &  21.6530$\pm$ 4.1083 &  \text{-}  &  36.1819$\pm$6.3786 &  \text{-}\\[1ex]
\cline{2-8}                 						
& $\lambda_2$ & 0.5865 $\pm$ 0.0051 & \text{-} &  0.5694$\pm$ 0.0050 &  \text{-} &   0.5594$\pm$0.0155 &  \text{-} \\ \cline{1-8}

\multirow{5}{*}{GSA}& $W\left(G\right)$ &  25.7549  $\pm$ 0.9448 & \textless 0.0001 &  54.3710$\pm$ 1.6604 & \textless 0.0001 &  78.9177$\pm$ 4.4068 & \textless 0.0001\\[1ex] % inserting body of the table			
\cline{2-8}
& $\mu\left(G\right)$ &  11.9479  $\pm$ 0.7929 & \textless 0.0001 &  9.9510$\pm$0.6087 & \textless 0.0001 & 6.0461 $\pm$0.9010 & \textless 0.0001\\[1ex]   	   	   	   	   	   	   				 \cline{2-8}
& $E_p$ &  0.0498  $\pm$ 0.0328 & \textless 0.0001 &  0.0498$\pm$0.0702 &  \textless 0.0001 &  0.6118$\pm$0.0560 & \textless 0.0001\\[1ex]	  	  	  	  							 \cline{2-8}
& $T_R$ &  1.0029  $\pm$ 0.3826 & \textless 0.0001 &  3.9853$\pm$0.6537 & \textless 0.0001 &  15.7807$\pm$3.3674 & \textless 0.0001\\[1ex]
\cline{2-8} 									
& $\lambda_2$ &  0.5973  $\pm$ 0.0282 & 0.1959 &  0.5634$\pm$0.0360 &  0.6271 &  0.5396$\pm$0.0409 &  0.1663\\[1ex]											
\hline\hline %inserts double horizontal lines
\end{tabular}
\label{table:result2a} % is used to refer this table in the text
\end{table*}

\begin{figure}[!h]
  \centering
  \includegraphics[scale=0.35]{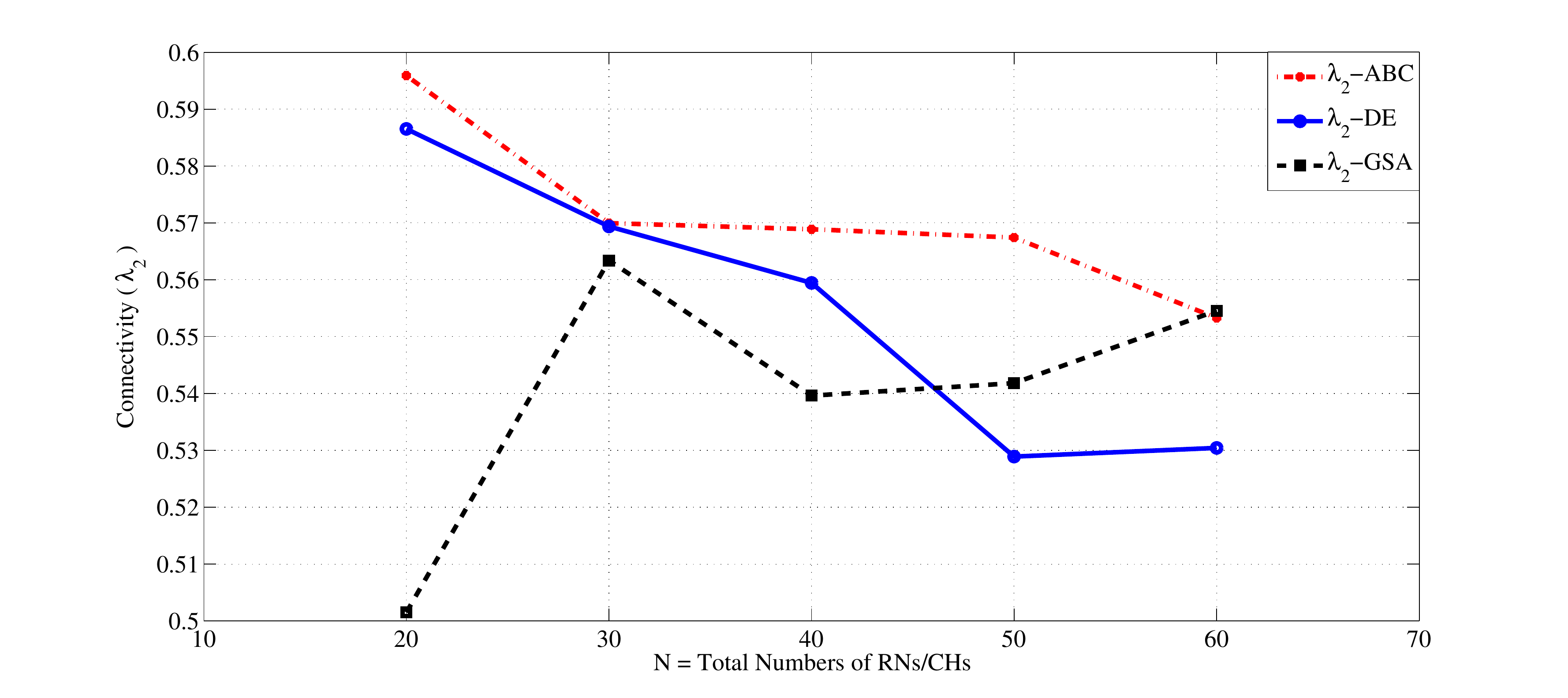}
  \caption{Connectivity vs. network size}\label{Fig_9_Lambda}
\end{figure}

\begin{figure}[!h]
  \centering
  \includegraphics[scale=0.35]{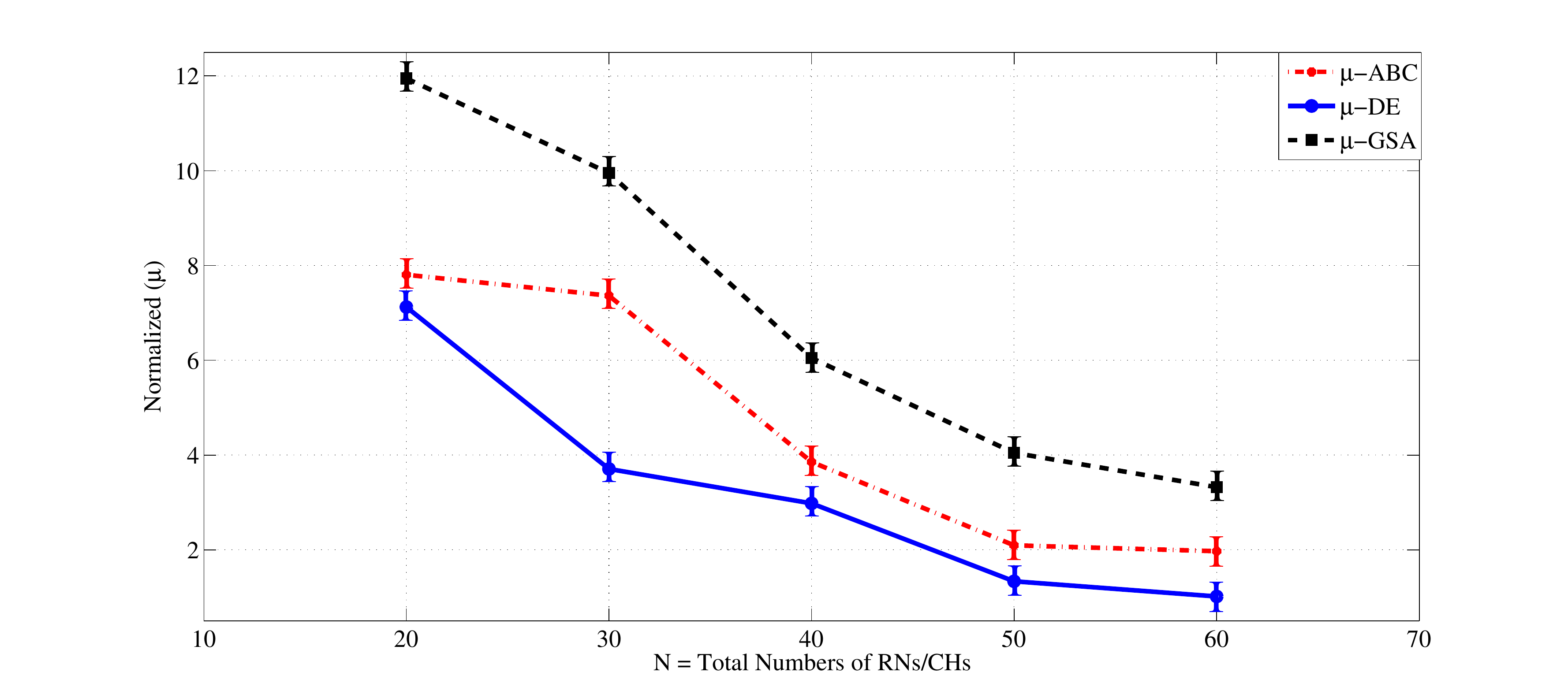}
  \caption{Normalized average distance vs number of nodes}\label{Fig_6_mu}
\end{figure}

\begin{figure}[!h]
  \centering
  \includegraphics[scale=0.35]{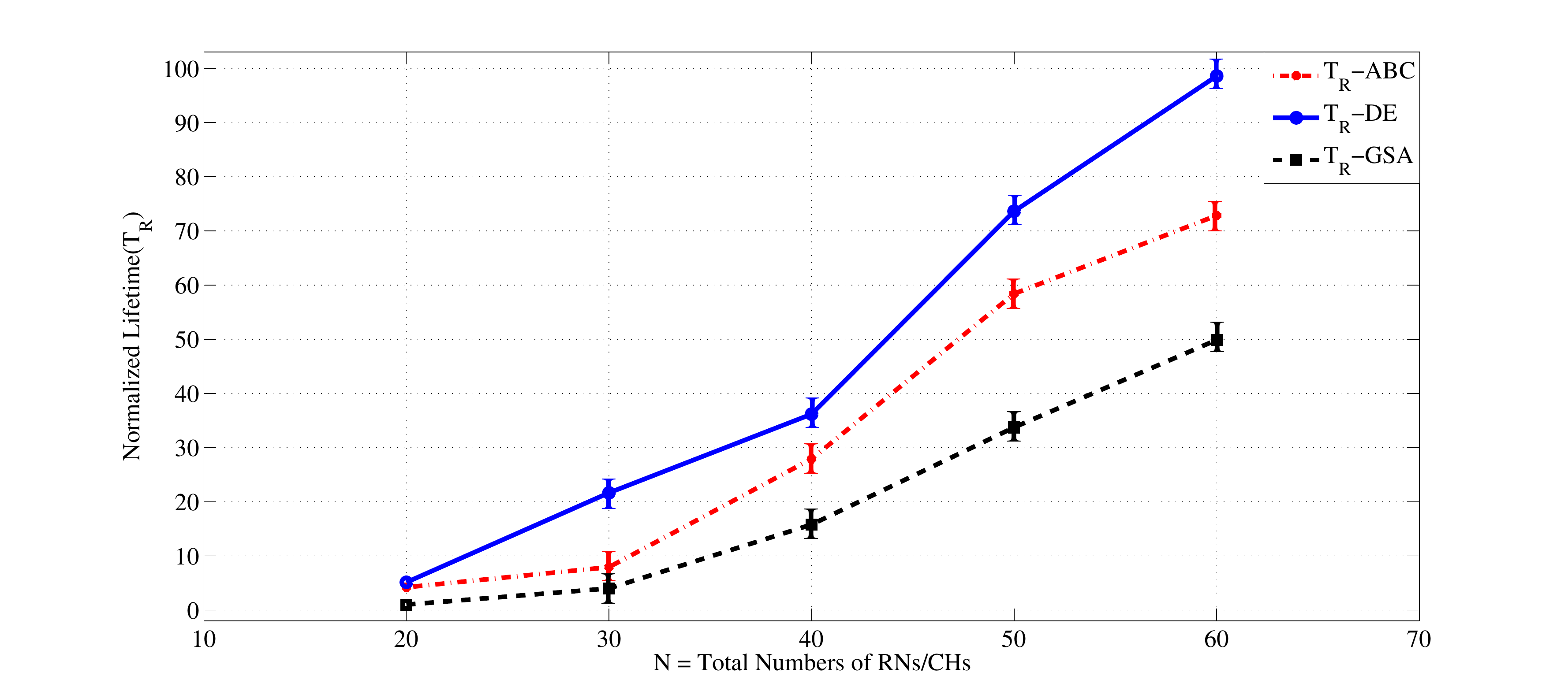}
  \caption{Network Lifetime against size with traffic level = 30 packets per round}\label{Fig_8_TR}
\end{figure}

\begin{figure}[!h]
  \centering
  \includegraphics[scale=0.35]{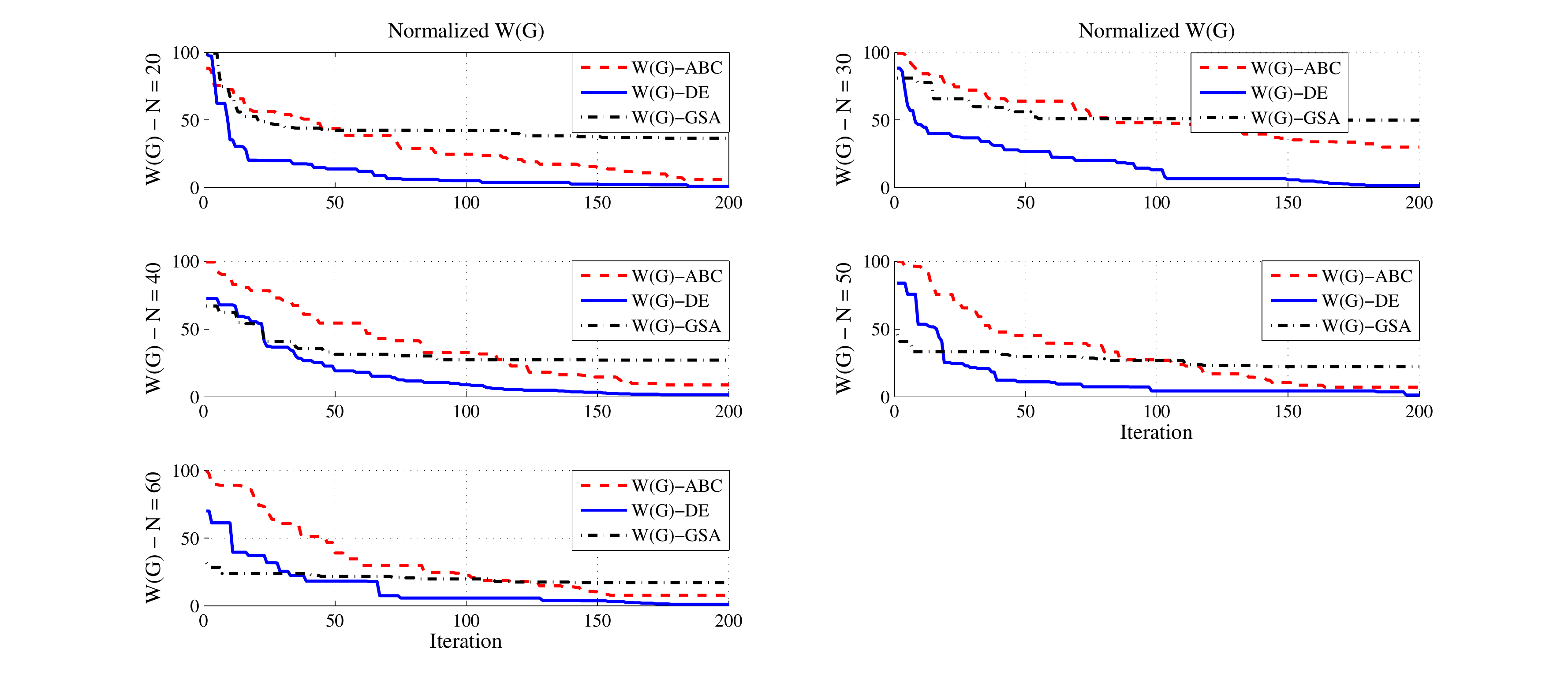}
  \caption{Comparison between objectives $W\left(G\right)$}\label{Fig_10_WG}
\end{figure}

\begin{table*}[!t]
 \setlength{\tabcolsep}{1pt}
\caption{Performance for network size $N=50$ and $60$} % title of Table
\centering % used for centering table
\small
\begin{tabular}{|c|c|c|c|c|c|} % centered columns (3 columns)
\hline %inserts double horizontal lines
\multicolumn{2}{|c|}{} & \multicolumn{2}{c|}{$N=50$} & \multicolumn{2}{c|}{$N=60$}\\ [0.5ex]
\cline{3-6}
\multicolumn{2}{|c|}{} &  Mean $\pm$ STD & \textit{p}-value &  Mean $\pm$ STD & \textit{p}-value \\ [0.5ex]
%heading
\hline\hline %inserts double horizontal lines        	   	   				
\multirow{5}{*}{ABC} & $W\left(G\right)$ & 93.9531$\pm$ 2.2281 & \textless 0.001 & 134.6739$\pm$ 1.8609 & \textless 0.001 \\[1ex] % inserting body of the table			
\cline{2-6}
& $\mu\left(G\right)$ &  2.0932$\pm$0.2900 & \textless 0.001  &  1.9695$\pm$0.1677 & \textless 0.001\\[1ex]  \cline{2-6}
& $E_p$ & 0.7755$\pm$ 0.0075 & \textless 0.001 &  0.7788$\pm$ 0.0042 & \textless 0.001\\[1ex]	 \cline{2-6}
& $T_R$ & 58.3868$\pm$5.0666 & \textless 0.001   &   72.8436$\pm$ 3.6710 & \textless 0.001 \\[1ex]  \cline{2-6} 		
& $\lambda_2$ & 0.5674$\pm$0.0089 & 0.001 &  0.5533$\pm$0.0131 & 0.0358\\ \cline{1-6}

\multirow{5}{*}{DE}& $W\left(G\right)$ & 88.1313$\pm$ 1.8857 & \text{-} &  124.0911$\pm$ 4.6739 & \text{-} \\[1ex] % inserting body of the table			
\cline{2-6}
& $\mu\left(G\right)$ & 1.3353$\pm$0.2455	& \text{-} & 1.0160$\pm$0.4211 & \text{-}	\\[1ex]  \cline{2-6}
& $E_p$ & 0.7935$\pm$0.0052 & \text{-} &   0.7998$\pm$0.0088 & \text{-}\\[1ex]	  	  	  	  							 \cline{2-6}
& $T_R$ & 73.6215$\pm$5.7131 & \text{-} &  98.6167$\pm$12.4156 & \text{-} \\[1ex]
\cline{2-6}                 						
&$\lambda_2$ & 0.5289$\pm$ 0.0155 & \text{-} &  0.5304 $\pm$0.0220 & \text{-}\\ \cline{1-6}

\multirow{5}{*}{GSA}& $W\left(G\right)$ & 108.9493$\pm$4.6732 & \textless 0.0001 &   149.6877 $\pm$6.7578 & \textless 0.001 \\[1ex] % inserting body of the table			
\cline{2-6}
& $\mu\left(G\right)$ & 4.0454$\pm$0.6083 & \textless 0.0001 & 3.3221 $\pm$ 0.6089	& \textless 0.001 \\[1ex]   	   	   	   	   	   	   				 \cline{2-6}
& $E_p$ & 0.7116$\pm$0.0236 & \textless 0.0001 & 0.7382$\pm$ 0.0196 & \textless 0.001  \\[1ex]	  \cline{2-6}
& $T_R$ & 33.7285$\pm$5.8655 & \textless 0.0001 &  49.8861$\pm$9.4138 & \textless 0.0001 \\[1ex]
\cline{2-6} 									
& $\lambda_2$ & 0.5418$\pm$0.0252 & 0.1526 &  0.5545$\pm$0.0361 &  0.1934 \\[1ex]											
\hline\hline %inserts double horizontal lines
\end{tabular}
\label{table:result2b} % is used to refer this table in the text
\end{table*}
In the first set of experiments, the network size is 20 (1 BS, 9 CHs, and 10 RNs) and it is incremented by 10 up to 60 for subsequent experiments.
The result in Fig.~\ref{Fig_9_Lambda} shows that overall network connectivity was preserved within desired range ($0.5$-$0.6$) specified by MST in Algorithm~\ref{alg:MST}. Fig.~\ref{Fig_6_mu} shows that all the techniques considered successfully minimize the average internode distance. However, it was observed that ABC-based deployment does not significantly improve as network size grows from 50 to 60. It is clear that DE-based deployment outperforms both ABC and GSA-based deployment in extending the useful network lifetime as vividly shown in Fig.~\ref{Fig_8_TR}. When the packets per round is set to 30 and the network size is increased to 60, DE-based deployment achieves almost twice as GSA, and around $30\%$ more than the ABC-based algorithm. Since network lifetime theoretically depends on the average inter-node distance as given in Eq. \eqref{MyEq14}, it came as no surprise that all the three heuristic-based deployments extend the network lifetime considerably. The objective function values have been averaged over eight experiments and plotted against the number of iterations for visualization. As observed in Fig. \ref{Fig_10_WG}, all heuristics minimize the cost function as expected. However, in most of the experiments, GSA got trapped in local minimum in less than 100 iterations, which is an indication of poor performance in comparison with DE and ABC. \\*[-.1pc]



\begin{figure}[!h]
  \centering
  \includegraphics[scale=0.35]{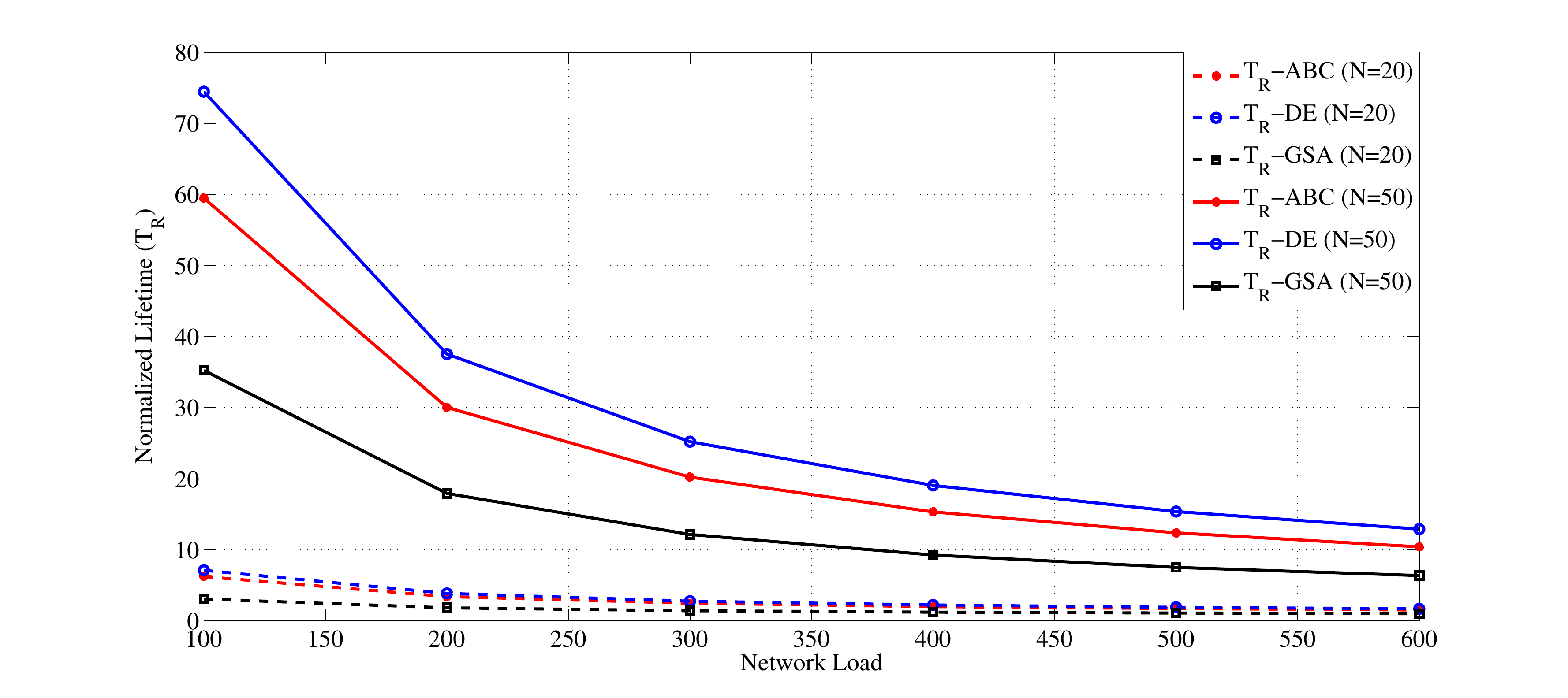}
  \caption{Lifetime vs. Network load of ABC, GSA and DE}\label{Fig_11_TR_ABC}
\end{figure}

\begin{figure}[!h]
  \centering
  \includegraphics[scale=0.35]{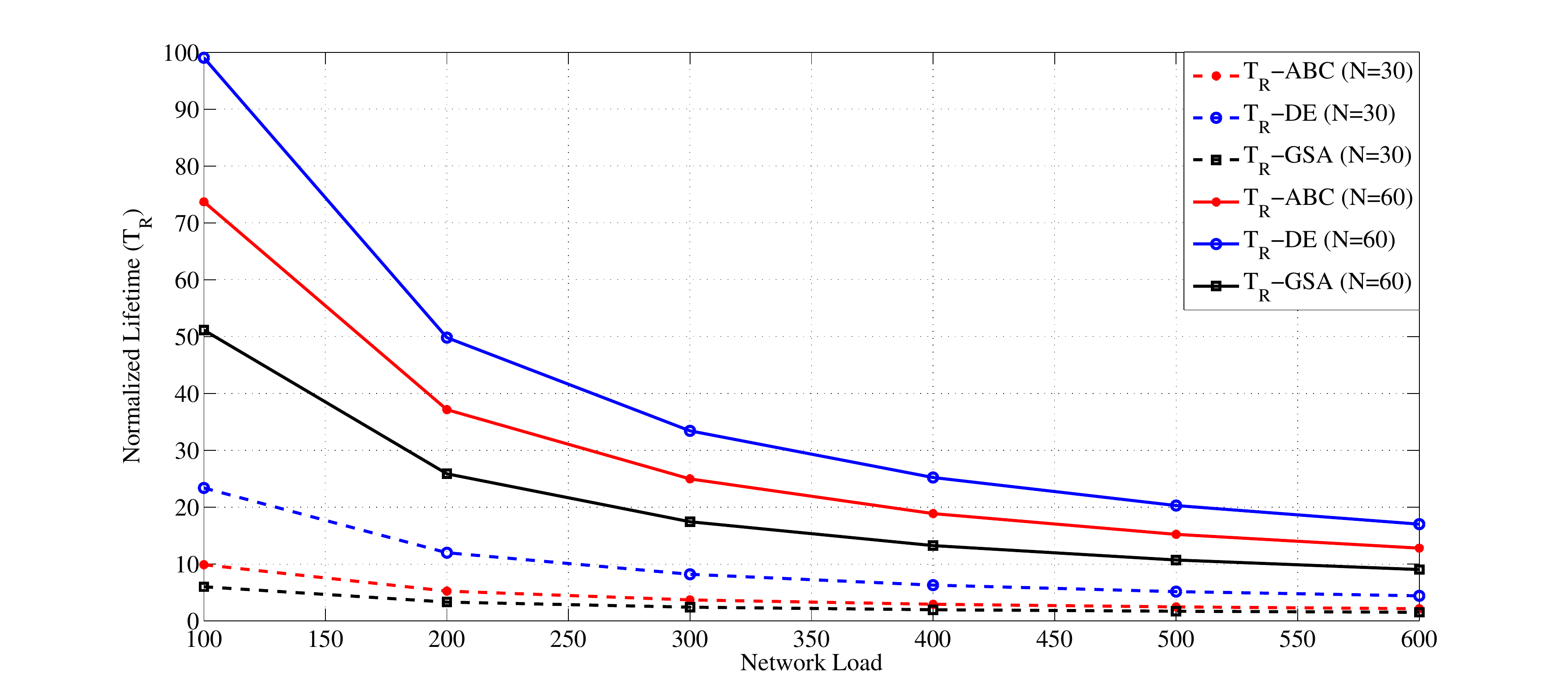}
  \caption{Lifetime vs. Network load of ABC, GSA and DE}\label{Fig_12_TR_DE}
\end{figure}

\begin{figure}[!h]
  \centering
  \includegraphics[scale=0.35]{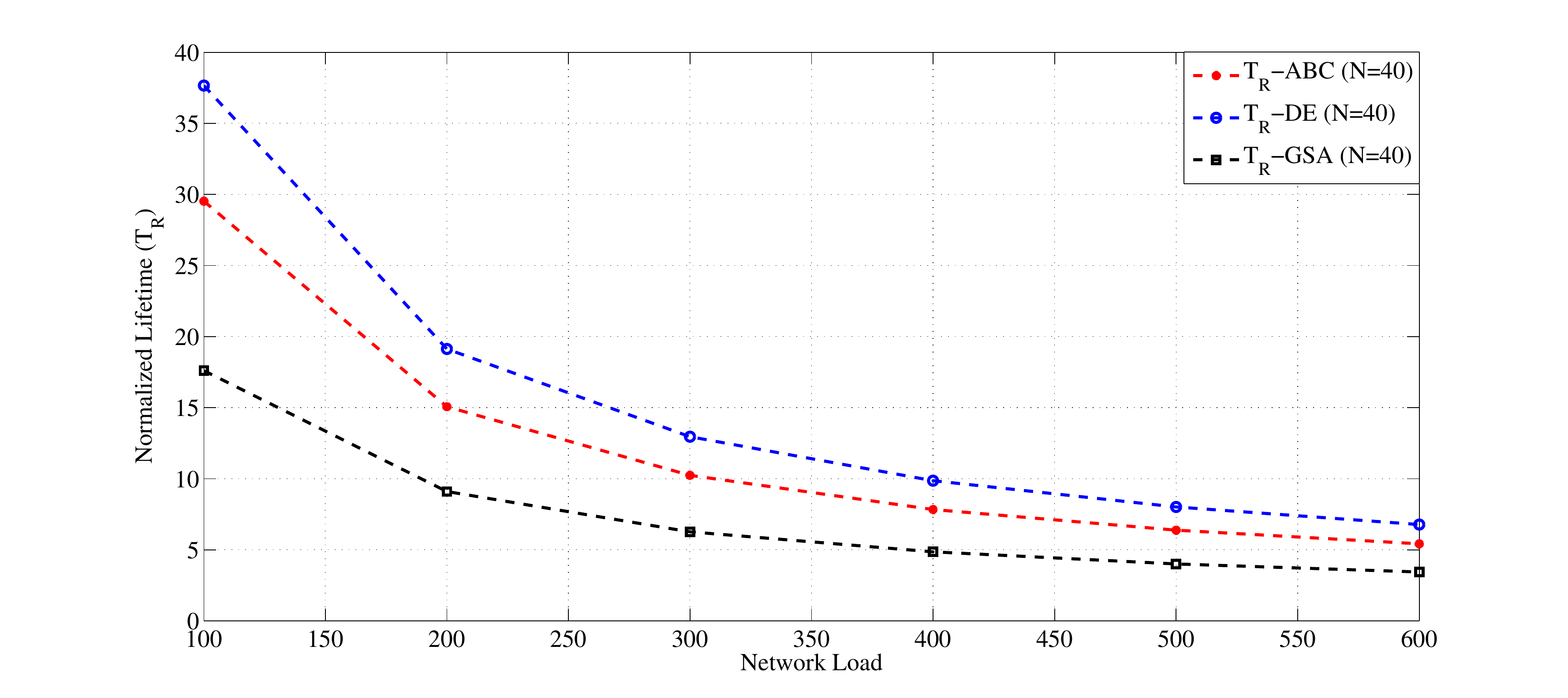}
  \caption{Lifetime vs. Network load of ABC, GSA and DE}\label{Fig_13_TR_GSA}
\end{figure}
The scenario where more traffics are generated due to more deployed sensor nodes or increased data collection rate is also investigated. In this scenario, CHs/RNs have more data to relay back and forth among themselves and to BS. This increased traffic could arise as a result of extra package generated by new sensors. When network traffic was varied, an increase in energy usage for packets transmission and reception was observed. By proportion, a drastic drop in useful network lifetime was observed for all the three heuristic-based deployment strategies, as shown in Figs.~\ref{Fig_11_TR_ABC}, \ref{Fig_12_TR_DE}, \ref{Fig_13_TR_GSA}. For DE-based algorithm, lifetime of approximately 17 rounds was obtained when the traffic was increased to 600ppr and network size to 60. On the other hand, ABC and GSA-based deployments only manage to achieve 13 and 8 rounds, respectively, with the same network configuration.\\*[.4pc]
\indent
Close scrutiny of Tables (\ref{table:result2a}) and (\ref{table:result2b}) also reveals that, on average, the values observed in eight experiments are close to that observed in each individual experiment. This is an indication that the proposed deployment strategies are robust to different initial populations. In fact, it can be inferred from the p-value in Table~(\ref{table:result2a}) that there is no significant difference between ABC and DE-based deployment for small-sized network ($N=20$). However, DE-based deployment is superior to that of GSA-based heuristic as inferred from the p-values. For large enough network ($N=30$ upward), DE-based deployment shows a significant improved performance than its counterparts.

\section{CONCLUSION}
In this paper, two other evolutionary techniques - Gravitational Search Algorithm (GSA) and Differential Evolution (DE) are employed for RNs placement and qualitatively compared with existing placement solution that uses ABC. The proposed strategy has been developed and formulated as an optimization problem. The formulation stands on the premises that energy utilization of two communicating nodes depends not only on the distance and the traffic between them, but also on the complexity that arises in 3-D settings with huge search space and high computational cost. By using a two-tier two-phase architecture, the complexity of the deployment can be circumvented. MST is employed in the first phase to deploy the backbone devices with minimum connectivity. Then, DE and GSA are utilized to locate the optimal positions of second-phase relay nodes for lifetime maximization with guaranteed cost and connectivity constraints. It is concluded that DE is more efficient for relay node deployment in 3D settings than ABC and GSA as shown in the experimental results. In future, potential direction would be to investigate the effect of RN grid mobility on network lifetime enhancement and another interesting extension would be to investigate the effect of interference due to RN's wider range on network throughput.

\section*{Appendix}
\setcounter{table}{0}
\renewcommand{\thetable}{A\arabic{table}}

\begin{table*}[!t]
 \setlength{\tabcolsep}{1pt}
\caption{Experimental results for network size $N=20$} % title of Table
\centering % used for centering table
\small
\begin{tabular}{|c| c| c| c| c|c| c| c| c|c|} % centered columns (3 columns)
\hline
\multicolumn{2}{|c|}{Experiment Index} &   1   &     2    &   3    &   4  &   5   &   6   &   7  &   8  \\ [0.5ex]
%heading
\hline\hline         	   	   				
\multirow{5}{*}{ABC} & $W\left(G\right)$ &   20.4779    & 20.8481   &  22.2951   &  20.5103   &  20.0171    & 20.8472   &  21.0671   &  20.4845 \\[1ex] 			 \cline{2-10}
& $\mu\left(G\right)$ &  7.5188  &    7.8295   &   9.0440    &  7.5460   &   7.1320   &   7.8288    &  8.0133   &   7.5244 \\[1ex]   	   	   	   	   	   	   				 \cline{2-10}
& $E_p$ & 0.5171    &   0.4927    &   0.3842     &  0.5150     &  0.5456   &    0.4928    &   0.4777   &    0.5166  \\[1ex]	  	  	  	  							 \cline{2-10}
& $T_R$ &  4.5307   &   4.1338   &   2.8554   &   4.4947   &   5.0730  &    4.1346   &   3.9138   &   4.5233 \\[1ex]
\cline{2-10} 									
& $\lambda_2$ &  0.5978   &     0.5958   &   0.5833    &  0.5968  &  0.5985   &  0.5981   &  0.5995 &  0.5973 \\\cline{1-10}

\multirow{5}{*}{DE}& $W\left(G\right)$ &   20.0751  &  19.8563   & 19.3830  &  20.9926   & 20.2172   & 20.0487  &  19.4318  &  20.0423 \\[1ex] 			 \cline{2-10}
& $\mu\left(G\right)$ &  7.1807   &  6.9971   &  6.5998   &  7.9508   &  7.3000  &   7.1586   &  6.6408   &  7.1532 	 \\[1ex]   	   	   	   	   	   	   				 \cline{2-10}
& $E_p$ & 0.5421  &  0.5551 &   0.5817 &   0.4828 &   0.5334  &  0.5437 &   0.5791  &  0.5441  \\[1ex]	  	  	  	  							 \cline{2-10}
& $T_R$ &  5.0016   &    5.2759   &   5.9186    &   3.9874  &   4.8306   &   5.0340   &  5.8490  &   5.0418  \\[1ex]
\cline{2-10}                 						
& $\lambda_2$ &  0.5871   &     0.5922   &   0.5833    &    0.5763    &   0.5912   &  0.5834 &   0.5865  &  0.5891  \\ \cline{1-10}

\multirow{5}{*}{GSA}& $W\left(G\right)$ &   26.5846   &  26.3823  &   24.2251  &   24.7767  &   26.5787  &   26.1230  &   24.9617   &  26.4071   \\[1ex] 			
\cline{2-10}
& $\mu\left(G\right)$ &  12.6443   &  12.4745  &   10.6639  &   11.1269  &   12.6393   &  12.2568   &  11.2821   &  12.4953  \\[1ex]   	   	   	   	   	   	   				 \cline{2-10}
& $E_p$ & 0.0910  &  0.0624   &  0.0020   &  0.0410  &  0.0901  &  0.0268   & 0.0195   & 0.0659  \\[1ex]	  	  	  	  							 \cline{2-10}
& $T_R$ &  0.6799   &  0.7475   &  1.6517  &   1.3842   &  0.6818  &   0.8379   &  1.3007   &  0.7391 \\[1ex]
\cline{2-10} 									
& $\lambda_2$ &  0.5621   &   0.5982  &    0.5952   &   0.5122     &    0.5743     &  0.5908   &   0.5887   &   0.5648 \\[1ex]											 
\hline\hline
\end{tabular}
\label{table:appen1}
\end{table*}
\begin{table*}[!t]
 \setlength{\tabcolsep}{1pt}
\caption{Experimental results for network size $N=30$} % title of Table
\centering % used for centering table
\small
\begin{tabular}{|c| c| c| c| c|c| c| c| c|c|}
\hline
\multicolumn{2}{|c|}{Experiment Index} &   1   &     2    &   3    &   4  &   5   &   6   &   7  &   8  \\ [0.5ex]
%heading
\hline\hline        	   	   				
\multirow{5}{*}{ABC} & $W\left(G\right)$ &   47.1533   &  49.4577   &  48.9101  &   46.0614  &   47.8593   &  46.2467  &   46.8965  &   45.8660  \\[1ex] 			
 \cline{2-10} 								
 & $\mu\left(G\right)$ &  7.3050   &   8.1498   &   7.9490  &    6.9047  &    7.5639  &    6.9727     & 7.2109    &  6.8331 \\[1ex]   	   	   	   	   	   	   				 \cline{2-10}
 & $E_p$ &  0.5331   &   0.4662  &    0.4830   &   0.5614  &    0.5136  &    0.5568   &   0.5399  &   0.5663  \\[1ex]	  	  	  	  							  \cline{2-10}
 & $T_R$ &   7.9923   &    6.3928    &   6.7414    &   8.8858  &     7.4638   &   8.7273   &    8.1938   &    9.0563 \\[1ex] 																			  \cline{2-10}    								
 & $\lambda_2$ &  0.5746   &     0.5628   &   0.5750    &  0.5673  &  0.5680   &  0.5724   &  0.5636 &  0.5758 \\
 \cline{1-10}

\multirow{5}{*}{DE}& $W\left(G\right)$ &  38.3427  &   37.4194   &  38.1365   &  36.0939  &   34.2659  &   38.4854  &   39.7146  &   36.2412  \\[1ex] % inserting body of the table			
\cline{2-10}
& $\mu\left(G\right)$ &  4.0751  &    3.7366   &   3.9995  &    3.2507  &    2.5805  &    4.1274   &   4.5780  &    3.3047  \\[1ex]   	   	   	   	   	   	   				 \cline{2-10}
& $E_p$ & 0.7117  &   0.7247 &    0.7147 &    0.7417   &  0.7625 &    0.7096   &  0.6907  &   0.7399 \\[1ex]	  	  	  	  							 \cline{2-10}
& $T_R$ &     19.1790  &   21.1123  &   19.5928  &   24.2843   &  29.5895   &  18.8985   &  16.6617  &   23.9061  \\[1ex]
\cline{2-10}                 						
& $\lambda_2$ &  0.5691   &   0.5737   &   0.5695  &    0.5741  &    0.5590   &   0.5662   &   0.5734  &    0.5698  \\ \cline{1-10}

\multirow{5}{*}{GSA}& $W\left(G\right)$ & 52.2730   &  53.1068  &   56.8519  &   56.0118  &   52.8760  &    53.5062  &   55.4777  &   54.8648  \\[1ex] % inserting body of the table			
\cline{2-10}
& $\mu\left(G\right)$ & 9.1819   &   9.4875  &   10.8605  &   10.5525   &   9.4029    &  9.6340   &  10.3567   &  10.1320 \\[1ex]   	   	   	   	   	   	   				 \cline{2-10}
& $E_p$ & 0.3705   &   0.3389  &    0.1766  &    0.2161   &   0.3478   &   0.3232   &   0.2402   &   0.2670 \\[1ex]	  	  	  	  							 \cline{2-10}
& $T_R$ &   4.8574    &   4.4736   &    3.0590   &    3.3377   &    4.5770    &   4.2997  &     3.5256  &     3.7521 \\[1ex]
\cline{2-10} 									
& $\lambda_2$ & 0.5041   &   0.5858   &   0.5963   &   0.5930   &   0.5314  &    0.5306   &   0.5724   &   0.5934 \\[1ex]										 
\hline\hline
\end{tabular}
\label{table:appen2}
\end{table*}
\begin{table*}[!t]
 \setlength{\tabcolsep}{1pt}
\caption{Experimental results for network size $N=40$}
\centering
\small
\begin{tabular}{|c| c| c| c| c|c| c| c| c|c|c|c|}
\hline
\multicolumn{2}{|c|}{Experiment Index} &   1   &     2    &   3    &   4  &   5   &   6   &   7  &   8 \\ [0.5ex]
%heading
\hline\hline        	   	   				
\multirow{5}{*}{ABC} & $W\left(G\right)$ &  66.5124   &  72.8250   &  66.6371  &   67.4001   &  68.5336  &   66.9232  &   68.9392  &   67.8106  \\[1ex] 			 \cline{2-10} 								
 & $\mu\left(G\right)$ &  3.5099   &  4.8005    & 3.5354   &  3.6914   &  3.9231   &  3.5939   &  4.0060   &  3.7753 \\[1ex]   	   	   	   	   	   	   				 \cline{2-10}
 & $E_p$ &  0.7329   &   0.6807   &   0.7320   &   0.7263   &   0.7176   &   0.7299    &  0.7144   &   0.7232 \\[1ex]	  	  	  	  							  \cline{2-10}
 & $T_R$ &   30.5445  &   21.3943   &  30.3250   &  29.0207   &  27.1999   &  29.8281  &   26.5805   &  28.3456 \\[1ex] 																			  \cline{2-10}    								
 & $\lambda_2$ &  0.5626    &   0.5785    &   0.5635    &   0.5682  &     0.5734   &    0.5613   &    0.5753   &    0.5682 \\
 \cline{1-10}

\multirow{5}{*}{DE}& $W\left(G\right)$ &   66.5340   &  64.4359  &   61.4971  &   62.5587   &  63.0498   &  62.5162  &   60.2059   &  70.5927\\[1ex] % inserting body of the table			
\cline{2-10}
& $\mu\left(G\right)$ &   3.5143   &   3.0854  &    2.4845   &   2.7016  &    2.8020   &   2.6929  &    2.2205    &  4.3441  \\[1ex]   	   	   	   	   	   	   				 \cline{2-10}
& $E_p$ & 0.7327   &   0.7472    &  0.7652   &   0.7590   &   0.7560   &   0.7592  &    0.7724  &    0.7007 \\[1ex]	  	  	  	  							 \cline{2-10}
& $T_R$ &     30.5063   &  34.4838  &   41.1096   &  38.5585  &   37.4410  &   38.6570  &   44.4823  &   24.2163 \\[1ex]
\cline{2-10}                 						
& $\lambda_2$ &  0.5435  &    0.5583  &    0.5411    &  0.5621   &   0.5574    &  0.5731  &    0.5518  &    0.5882 \\ \cline{1-10}

\multirow{5}{*}{GSA}& $W\left(G\right)$ & 75.6850  &   76.7186   &  74.3766  &   76.7740   &  85.4740   &  85.4565   &  76.2813  &   80.5751  \\[1ex] % inserting body of the table			
\cline{2-10}
& $\mu\left(G\right)$ &  5.3852   &   5.5965   &   5.1177   &   5.6079   &   7.3865   &   7.3830   &   5.5071    &  6.3850 \\[1ex]   	   	   	   	   	   	   				 \cline{2-10}
& $E_p$ & 0.6522   &   0.6410   &   0.6656   &   0.6404    &  0.5270  &    0.5273   &   0.6458    &  0.5954 \\[1ex]	  	  	  	  							 \cline{2-10}
& $T_R$ &   18.3052  &   17.3143   &  19.6516  &   17.2629  &   10.9339   &  10.9438 &    17.7260   &  14.1080 \\[1ex]
\cline{2-10} 									
& $\lambda_2$ &  0.5810   &   0.5241   &   0.4720  &    0.5104  &    0.5144   &   0.5568  &    0.5671   &   0.5910 \\[1ex]									 \hline\hline
\end{tabular}
\label{table:appen3}
\end{table*}
\begin{table*}[!t]
 \setlength{\tabcolsep}{1pt}
\caption{Experimental results for network size $N=50$}
\centering
\small
\begin{tabular}{|c| c| c| c| c|c| c| c| c|c|c|c|}
\hline
\multicolumn{2}{|c|}{Experiment Index} &   1   &     2    &   3    &   4  &   5   &   6   &   7  &   8  \\ [0.5ex]
\hline\hline         	   	   				
\multirow{5}{*}{ABC} & $W\left(G\right)$ &  91.0270    & 94.8536   &  91.5182  &   96.3813  &   97.2481 &    92.3952  &   93.7660  &   94.4354 \\[1ex] 			 
 \cline{2-10} 								
 & $\mu\left(G\right)$ &  1.7123  &    2.2104  &    1.7762  &    2.4093   &   2.5221   &   1.8904   &   2.0688    &  2.1560 \\[1ex]   	   	   	   	   	   	   				 \cline{2-10}
 & $E_p$ &  0.7851   &   0.7727    &  0.7836    &  0.7673   &   0.7642   &   0.7809   &   0.7764   &   0.7741  \\[1ex]	  	  	  	  							  \cline{2-10}
 & $T_R$ &   65.2941 &    56.1513  &   64.0273   &  52.9281   &  51.1961   &  61.8367  &   58.5872  &   57.0735 \\[1ex] 																			  \cline{2-10}    								
 & $\lambda_2$ &  0.5776   &   0.5638  &    0.5761   &   0.5653   &   0.5510  &    0.5649 &     0.5643  &    0.5762 \\
 \cline{1-10}
\multirow{5}{*}{DE}& $W\left(G\right)$ &   90.4224   &  89.7886   &  85.7999  &   87.5349   &  90.0820   &  88.0340  &   87.9512   &  85.4375 \\[1ex] 		 \cline{2-10}
& $\mu\left(G\right)$ &    1.6335   &   1.5510   &   1.0318  &    1.2577   &   1.5892   &   1.3226  &   1.3118   &   0.9846   \\[1ex]   	   	   	   	   	   	   				 \cline{2-10}
& $E_p$ & 0.7870    &  0.7889  &    0.7999  &    0.7953  &    0.7880  &    0.7939  &    0.7941   &   0.8008  \\[1ex]	  	  	  	  							 \cline{2-10}
& $T_R$ &  66.8945  &   68.6220  &   80.7908  &   75.2074  &   67.8159  &   73.6867  &   73.9365  &   82.0185 \\[1ex]
\cline{2-10}                 						
& $\lambda_2$ &  0.5352   &   0.5035   &   0.5132  &    0.5435   &   0.5412   &   0.5239    &  0.5471  &     0.5235 \\ \cline{1-10}
\multirow{5}{*}{GSA}& $W\left(G\right)$ & 105.1462 &  113.0629 &  108.1924 &  115.8275 &  110.3703 &  100.5047 &  109.0259 &  109.4644 \\[1ex] 			 \cline{2-10}
& $\mu\left(G\right)$ &  3.5503   &   4.5809   &   3.9468   &   4.9408  &    4.2303  &    2.9461  &    4.0553    &  4.1124  \\[1ex]   	   	   	   	   	   	   				 \cline{2-10}
& $E_p$ &  0.7314    &  0.6905  &    0.7167   &   0.6741    &  0.7054   &   0.7516   &   0.7125   &   0.7102 \\[1ex]	  \cline{2-10}
& $T_R$ &   38.1256   &  28.7571  &   34.1548   &  26.1324 &    31.6085  &   45.2562  &   33.1534  &   32.6400  \\[1ex]
\cline{2-10} 									
& $\lambda_2$ &  0.5283  &   0.5304   &   0.5470   &  0.5490    &   0.5269    &   0.5208    &   0.5992     &  0.5328 \\[1ex]										 \hline\hline
\end{tabular}
\label{table:appen4}
\end{table*}
\begin{table*}[H]
\setlength{\tabcolsep}{1pt}
\caption{Experimental results for network size $N=60$}
\centering
\small
\begin{tabular}{|c| c| c| c| c|c| c| c| c|c|}
\hline
\multicolumn{2}{|c|}{Experiment Index} &   1   &     2    &   3    &   4  &   5   &   6   &   7  &   8 \\ [0.5ex]
\hline\hline         	   	   				
\multirow{5}{*}{ABC} & $W\left(G\right)$ & 135.6958  & 133.2450  & 131.7722  & 137.7684  & 134.7591  & 133.6464  & 134.4074  & 136.0973 \\[1ex] 		 \cline{2-10} 								
 & $\mu\left(G\right)$ &  2.0615    &   1.8407    &   1.7080   &    2.2483    &   1.9771   &    1.8769     &  1.9455   &    2.0977	\\[1ex]   	   	   	   	   	   	   				 \cline{2-10}
 & $E_p$ &   0.7766   &   0.7821  &    0.7852  &    0.7717  &    0.7787   &   0.7812  &    0.7795   &   0.7757 \\[1ex]	  	  	  	  							  \cline{2-10}
 & $T_R$ &   70.7615   &  75.6364  &   78.7573   &  66.9271   &  72.5796   &  74.8114  &   73.2763   &  69.9989  \\[1ex] 																			  \cline{2-10}    								
 & $\lambda_2$ &  0.5542   &   0.5352  &    0.5486   &   0.5671  &    0.5724  &    0.5596  &    0.5371  &    0.5519 \\
 \cline{1-10}

\multirow{5}{*}{DE}& $W\left(G\right)$ &   119.8184  &  122.0953  &  119.5098 &   128.1753   & 123.5144  &  133.0943  &  125.5230 &   120.9982 \\[1ex] % inserting body of the table			
\cline{2-10}
& $\mu\left(G\right)$ &    0.6311   &   0.8362   &   0.6033   &   1.3840   &   0.9640   &   1.8272   &   1.1450    &  0.7374	\\[1ex]   	   	   	   	   	   	   				 \cline{2-10}
& $E_p$ & 0.8075    &   0.8037  &     0.8080   &    0.7926   &    0.8012 &      0.7824    &   0.7976   &   0.8056  \\[1ex]	  	  	  	  							 \cline{2-10}
& $T_R$ &  110.6521  &  103.5356  &  111.6609   &  87.0439  &   99.3756   &  75.9489   &  93.8222 &   106.8941 \\[1ex]
\cline{2-10}                 						
& $\lambda_2$ &  0.5003    &  0.5380   &   0.5437   &   0.5605  &    0.5434  &    0.5051  &    0.5418    &  0.5105 \\ \cline{1-10}

\multirow{5}{*}{GSA}& $W\left(G\right)$ & 150.9769  &  135.9783   & 154.0231  &  146.3812  &  153.6548   & 146.3354  &  152.3104  &  157.8413  \\[1ex] % inserting body of the table			
\cline{2-10}
& $\mu\left(G\right)$ & 3.4383   &   2.0870   &   3.7128  &    3.0243  &    3.6796  &    3.0201   &   3.5584   &   4.0568 \\[1ex]   	   	   	   	   	   	   				 \cline{2-10}
& $E_p$ &  0.7354    &   0.7759   &    0.7255    &   0.7491  &     0.7268   &    0.7492   &    0.7311   &    0.7124  \\[1ex]	  	  	  	  							 \cline{2-10}
& $T_R$ &   47.5134   &  70.2239  &   44.0275   &  53.4038    & 44.4326  &   53.4667   &  45.9491   &  40.0714 \\[1ex]
\cline{2-10} 									
& $\lambda_2$ &  0.5091   &   0.5757  &    0.5145  &    0.5246   &   0.5433   &   0.5919  &    0.5775    &  0.5997 \\[1ex]
\hline\hline
\end{tabular}
\label{table:appen5}
\end{table*}

\bookmarksetup{startatroot}
\bibliographystyle{IEEEtran}
\bibliography{sensorbibnew}
\end{document}